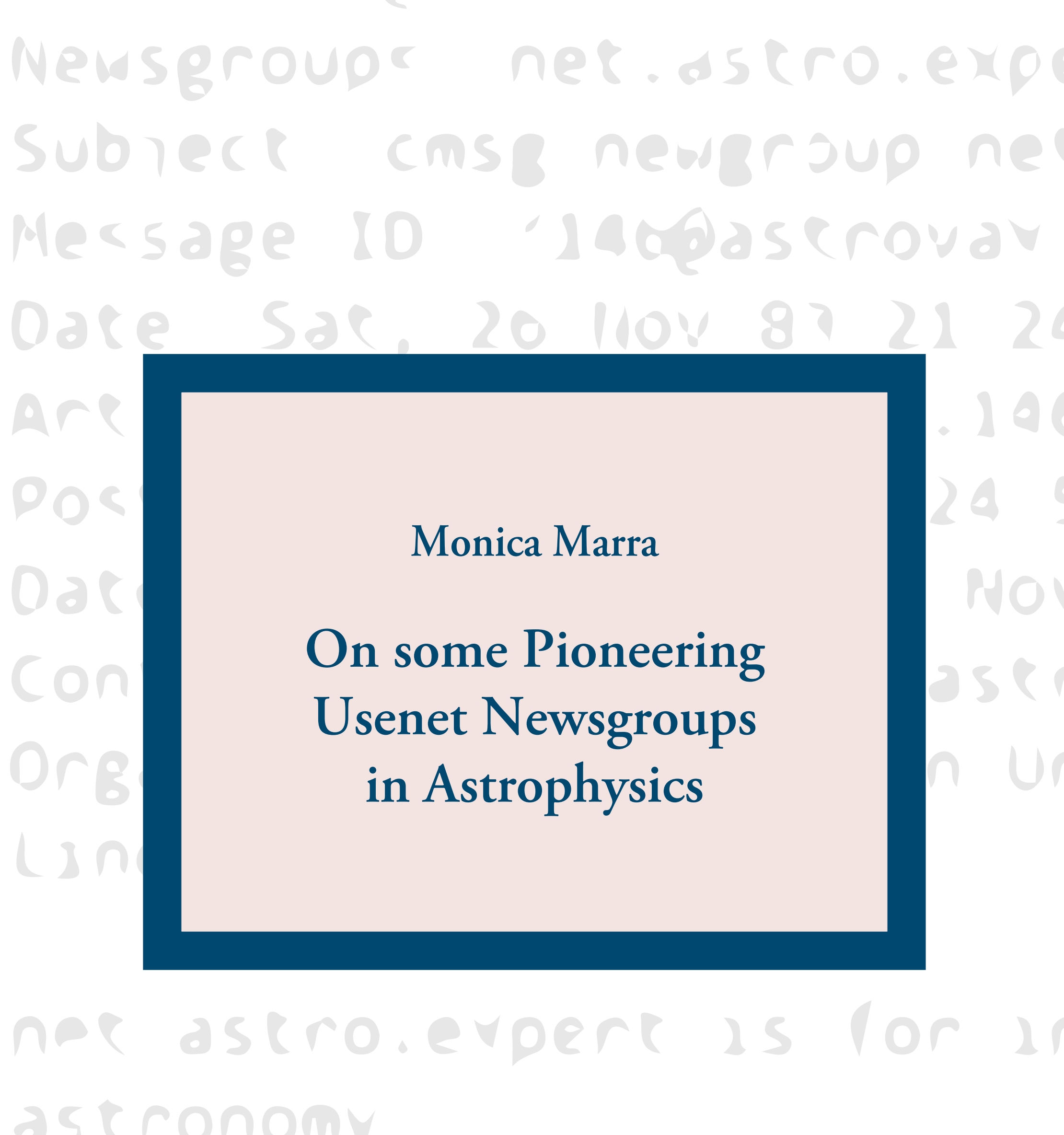

Monica Marra

# On some Pioneering Usenet Newsgroups in Astrophysics

Monica Marra

# On some Pioneering Usenet Newsgroups in Astrophysics

With a Preface by Dr. Phillip Helbig

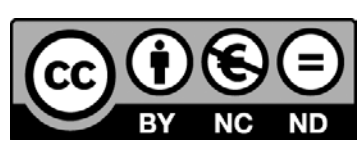

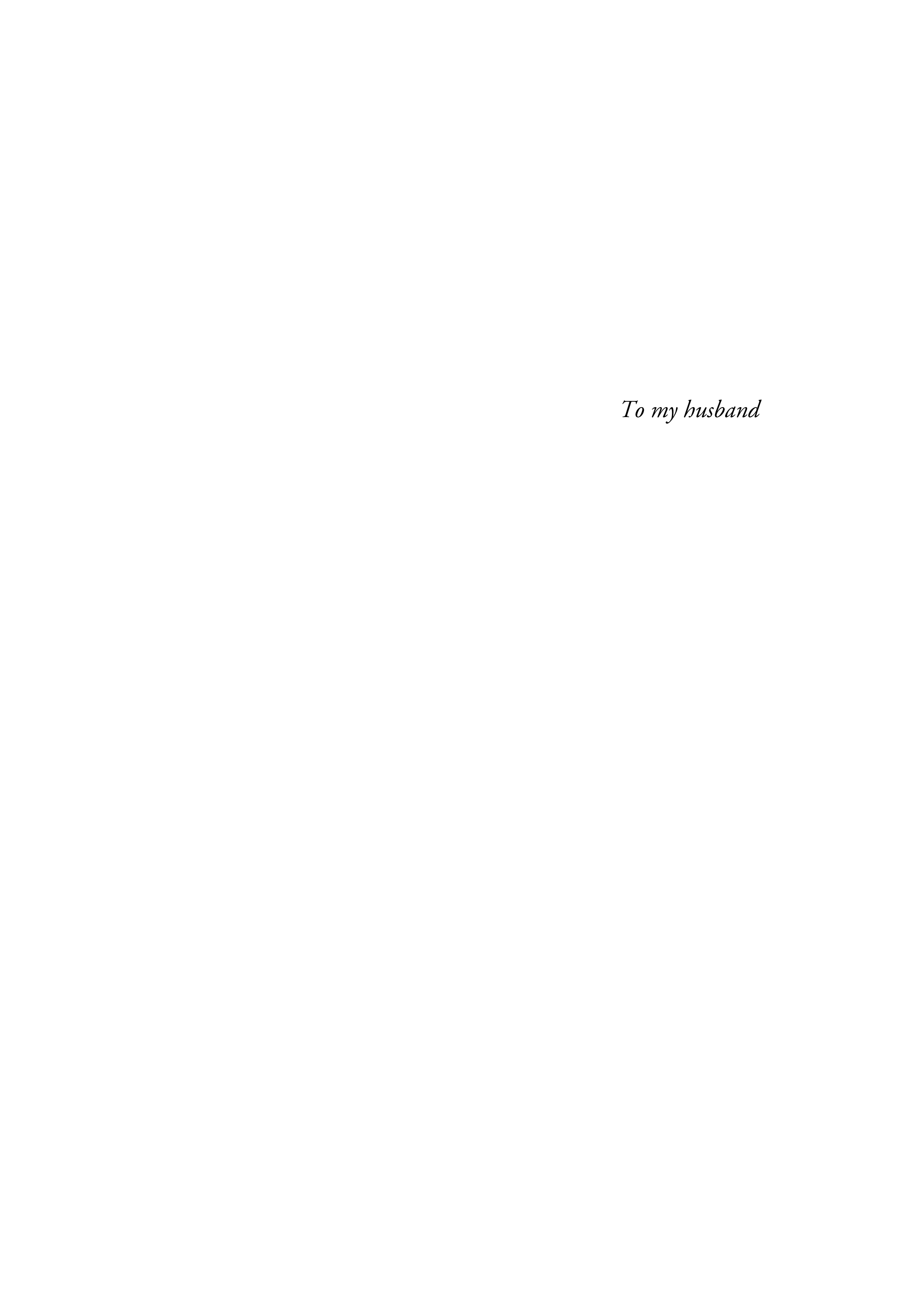

*To my husband*

# *Preface*

It's 2025 and someone is doing research on the history of Usenet. That in itself is a worthy goal, but perhaps it could also remind us of when, for many, the internet really took off and what has changed since then. Although the internet itself dates from the late 1980s (and its forerunner ARPANET to the late 1960s), and Usenet from the late 1970s, it was not until the early 1990s that many academics had access to the internet, which greatly increased the number of Usenet users, around the same time as the World Wide Web became publicly available. I remember a colleague who used to check all the websites in the world every morning to see what was new – there were thirteen of them. In contrast, there were already thousands of newsgroups. Although the fraction of internet traffic due to Usenet has decreased over time, in absolute terms the volume has steadily increased. While email was used for the same purpose as conventional mail, and FTP was a faster way to transfer files than magnetic tapes in a car, Usenet offered something qualitatively different: virtual communities built around a particular topic. Moderated groups offered an alternative for those who wanted a higher signal-to-noise ratio. As the fraction of academics on Usenet was higher then, in particular moderated groups on scientific topics arose, allowing discussion between amateurs and professionals in a particular field.

I started using Usenet in 1992, only shortly after gaining internet access via the university and also only shortly after having started working with computers. At the time, I was (and still am) writing Fortran code on computers running the VMS operating system, thus comp.lang.fortran and comp.os.vms were the most frequented groups by me: high volume, high signal-to-noise ratio, many professional participants (some of whom I later met in real life). As I was working in astrophysics (and still am), sci.physics.research and sci.astro.research were also on my list of subscribed newsgroups; I became a co-moderator of the former in 1997 and of the latter in 2016. The support for newsgroups has declined over the years; as a result, in 2016 I took over the moderation relays (automatically emailing submitted posts to moderated newsgroups to one of the active moderators, who if applicable will approve it and post it to the group) for those two newsgroups as well.

Today, the internet is ubiquitous, and most traffic is on the World Wide Web, which has moved beyond static web pages to include interactive sites (many of them commercial), forums, blogs, streaming services, and social media (the last two probably making up the bulk of traffic today). However, Usenet offers several advantages over many common internet applications: one newsreader can automatically move through a list of subscribed groups; one can use one's favourite editor for all posts; one can make entire threads invisible if they are not interesting. Also, since they are mostly plain text, posts are easily saved for future reference, and easy to archive. Newsgroups were dedicated to a particular topic, yet at the same time open to all; these days, most interaction on the internet takes place among small groups of people, or is open to the world but poorly organized, often with some secret algorithm determining what is seen rather than the users themselves. Perhaps most important was the sense of community – despite flames (some of which, however, were good-natured), most participants were there to help others rather than blow their own horn. It's still there for those who want it.

Monica Marra has provided a fascinating history of the early days (even before my time) of sci.astro. and sci.astro.research, which to a large extent mirrors the early history of other academic newsgroups. Some who took part in the foundation of Usenet are no longer with us, and others will leave us soon; it is thus good for posterity that she wrote this history while many of the sources are still available.

Dr. Phillip Helbig

# *Foreword*

The background to the present study was the idea of sketching some elements about a specific kind of blog, written by professional astrophysicists, which appeared to be meaningful from a socio-professional point of view.

In fact, though, it soon became clear that trying to add something to what was known of blogs in astrophysics without taking due account of the previous and/or contemporary online communication practices within this scholarly community would have been chimerical. This led to considering the early Internet as an appropriate study field.

Ideally, a comprehensive approach to this subject should include, at the very least, the adoption path of emails, mailing lists and newsgroups, early CMC tools which may be used formally, but also – certainly in the case of newsgroups – mainly informally. Seminal studies have demonstrated the central role of informal scholarly communication in reaching a crucial element for the advancement of science: consensus about new developments. This was proved for high energy physics[1], with whom we believe astrophysics shares some of its inner dynamics. To the historical significance of a reconstruction of this path – how the community created and uptook CMC tools –, research on this subject can add the feature of stressing the importance of informal communication in online settings in astrophysics, which seems to have been underexplored hitherto. A special room in a very specific context might be kept for ArXiv, the well-known online preprint repository that changed the scholarly communication in physics and related areas since its foundation in 1991 (its section for astrophysics, astro-ph, was born in April 1992), with a different approach. This territory being clearly too vast for a single piece of research, the present investigation will focus on Usenet newsgroups.

---

[1] Diana Crane, *Invisible Colleges: Diffusion of Knowledge in Scientific Communities* (University of Chicago Press, 1972), 64-65; Sharon Traweek, *Beamtimes and Lifetimes: The World of High Energy Physicists* (Harvard University Press, 1988), 117-18. In a specific perspective: Karin D. Knorr-Cetina, *Epistemic Cultures. How the Sciences Make Knowledge* (Harvard University Press, 1999), 173-78.

Specifically, the research object consists of two pioneering wide-area newsgroups dedicated to astrophysics. They were conceived, created and maintained by members of the community itself primarily as venues for lively discussions. Mentions of parallel experiences in the nearby scholarly communities of physicists and of mathematicians will be made only when strictly necessary, leaving a more integrated account of this subject to a possible future occasion.

A partial draft of this research, containing the reconstruction of the experiences examined, has been posted to Arxiv (astro-ph) on February 24, 2024[2] for checking it with the community's knowledge. No negative remarks have been received.

[2] Monica Marra, "New stairways to the stars. Birth and evolution of two pioneering Usenet newsgroups in astrophysics (1983-1994)", https://arxiv.org/abs/2402.15845.

## *Acknowledgements*

The author would like to thank for their invaluable input (in alphabetical order): Vincenzo Antonuccio-Delogu, Robert Hanisch, Phillip Helbig, Joseph Lazio and the researchers of the Italian National Institute for Astrophysics who answered her questionnaire. Any misunderstanding of their contribution is the author's responsibility.

## *List of abbreviations*

GG: Google Groups
CMC: computer-mediated communication
C1: individual astrophysicist #1, email correspondence with author, March 16, 2023.
C2: individual astrophysicist #2, email correspondence with author, January 31, 2022.
C3: individual astrophysicist #3, telephone interview by author, January 16, 2023.
C4: individual astrophysicist #4, email correspondence with author, February 20, 2023.

## 1. *Introduction*

The present research will reconstruct the creation and investigate the milestones of two very early Usenet newsgroups dedicated to astrophysics, in the first nine years of their activity: 1983-1994. Together with mailing lists, Usenet newsgroups at large have an acknowledged role in distributing valuable historical information about definite communities.[1] By considering, for as much as it has been done, how these newsgroups served the astrophysics community, the present research contributes knowledge of some social dynamics at play in this specific context by the end of the twentieth century, while adding new elements to this environment's uptake of new communication technology.[2] We also believe that creating these newsgroup primarily for communication rather than documentation purposes, as it will be conveyed, and the degree of usefulness these CMC tools reached in the community are elements to build upon for rebalancing, to some extent, the relative importance of informal communication in online settings within the discipline. Finally, although the preferred perspective was that of highlighting the dynamic aspects of these experiences, it is believed this research may help the preservation of digital heritage, as more than forty years have passed since the debut of these newsgroups and they may be at risk of being forgotten.

This study forced us to confront some issues of sources and methods in web history, and required to contextualize these experiences within some major steps of Usenet history. Additionally, the main features of some landmark social phenomena such as, e.g., the evolution of the internet, the changing nature of the scientific communication over time and the growing role of non-professionals for a part of the activities in astrophysics required to be kept in mind as a background. In the last paragraph, after briefly sketching some acknowledged social changes which affected astrophysics in the second half of last century, we'll explain that our approach for making sense of some crucial aspects of our research object is a grassroots perspective, the results of which seem anyway to be of some sociological significance.

[1] Alexandre Hocquet and Frédéric Wieber, "Mailing List Archives as Useful Primary Sources for Historians: Looking for Flame Wars," *Internet Histories* 2, no. 1-2 (2018): 39.

[2] "It is important to remember the role of agency in change: the important role of the reception […] and implementation of new policies and responses to changing environments by academic staff themselves." (Tony Becher and Paul Trowler, *Academic Tribes And Territories: Intellectual Enquiry and the Culture of Disciplines*. 2.ed. (The Society for Research into Higher Education and Open University Press, 2001), 16).

## 2. *Usenet Newsgroups, a major online resource until the advent of the World Wide Web*

Usenet was created in 1979 and its first newsgroups appeared at the beginning of 1980.[1]

An early, simple and – unsurprisingly – entirely North American definition of Usenet newsgroups can be found in an unsigned newsgroup post dated April 20, 1982:

> *"USENET is an international network of UNIX sites, with hookups into the ARPA network, too. It is basically a fancy electronic Bulletin Board system. Numerous BTL machines are connected* […] *In addition, there are major sites at universities: UC Berkeley, Duke, U Waterloo, and so on* […] *and at industry nationwide: DEC, Tektronics, Microsoft, Intel, etc. There are numerous bulletin board categories, set up in a hierarchy. The first "node" in a category name indicates the breadth of distribution, later nodes indicate content.* […] *Newsgroup naming conventions: NO prefix= LOCAL ONLY; btl. = Bell Labs; net. = USENET wide categories; fa. = from ARPA-Net (no return feed, except via mail)."*[2]

---

[1] Henry Edward Hardy, "The History of the Net" (Master's Thesis, Grand Valley State University, MI, USA, 1993), archived November 26, 2024, at https://web.archive.org/web/20241126145904/http://www.devin.com/cruft/hardy.html; Katharine Mieszkowski, "The Geeks Who Saved Usenet," *Salon*, January 8, 2002, archived June 20, 2022, at https://web.archive.org/web/20220620153342/https://www.salon.com/2002/01/08/saving_usenet/; Ronda Hauben, "Commodifying Usenet and the Usenet Archive or Continuing the Online Cooperative Usenet Culture?," *Science Studies* 15, no. 1 (2002): 64, https://citeseerx.ist.psu.edu/viewdoc/download;jsessionid=B2D0D90EFE821D600B4304A3C51C9157?doi=10.1.1.468.8369&rep=rep1&type=pdf, accessed November 24, 2024; Steve Bellovin, "The Early History of Usenet. Part VI: The Public Announcement," *CircleID*, November 27, 2019, archived January 23, 2024, at https://web.archive.org/web/20240123130936/https://circleid.com/posts/20191127_the_early_history_of_usenet_part_vi_the_public_announcement (Bellovin was a co-founder).

[2] "Newsinfo.shell", unsigned, April 20, 1982, archived January 1, 2024, at https://web.archive.org/web/20241201180856/https://www.usenetarchives.com/view.php?id=net.sources&mid=PGFuZXdzLkF1Y2JhcnBhLjExODI%2B.

In contrast to the marginal role Usenet newsgroups presently play,[3] Hyman conveniently reminds that "USENET was fabulously successful, growing very rapidly from a few computers in North Carolina, and soon spreading to hundreds of systems throughout the world, but predominantly in North America."[4] Also,

> *"Usenet [was featured] as a burgeoning locus of Internet culture, to the point where it became a metonym for "the Net" of the 1990s itself. For a generation of young college students, Usenet served as their first experience with the Internet* [...]. [...] *this period marked Usenet's peak use, followed by a gradual decline at the start of the new millennium* [...] *as its underlying communications protocols were superseded by the Internet's TCP/IP (Russell, 2014). Usenet was used not just for discussion, but also for producing and circulating a huge volume of informational, educational, humorous, and folkloric material, including technical standards, tutorials."*[5]

The two Usenet newsgroups examined are sci.astro (unmoderated) and sci.astro.research (moderated), together with the predecessors we'll see. According to the short scope notes made available since their debut, sci.astro was dedicated to "astronomy dis-cussions and information,”[6] while sci.astro.research was presented as a “forum in astronomy/astrophysics research”.[7] Their activity has been substantially uninterrupted since.

[3] E.g.: Camille Paloque-Bergès, "Usenet as a Web Archive: Multi-Layered Archives of Computer-Mediated Communication," in *Web 25: Histories from the First 25 Years of the World Wide Web*, ed. Niels Brugger (Lang, 2017), 232. Since February 22, 2024, Google "has stopped Google Groups (at groups.google.com) to post content to Usenet groups, subscribe to Usenet groups, or view new Usenet content" (Rob Pegoraro, "End of an Era: Google Groups to Drop Usenet Support," *PC Mag*, December 16, 2023, archived January 26, 2024, at https://web.archive.org/web/20241126143729/https://www.pcmag.com/news/end-of-an-era-google-groups-to-drop-usenet-support).

[4] Avi Hyman, "Twenty Years of ListServ as an Academic Tool," *The Internet and Higher Education* no. 6 (2003): 18-19.

[5] Tristan Miller, Camille Paloque-Bergès & Avery Dame-Griff, "Remembering Netizens: An Interview with Ronda Hauben, Co-Author of Netizens: On the History and Impact of Usenet and the Internet (1997)", *Internet Histories* 7, no.1 (2023), https://doi.org/10.1080/24701475.2022.2123120.

[6] “Sci.astro”, GG, archived January 23, 2024, at https://web.archive.org/web/20240123144439/https://groups.google.com/g/sci.astro/about.

[7] “Sci.astro.research”, GG, archived January 23, 2024, at https://web.archive.org/web/20240123144829/https://groups.google.com/g/sci.astro.research/about.

## 3. *Methodological foreword*

As it has been clarified for the Net at large,[1] trying to draw historical accounts of CMC tools' uptakes as linear worldwide processes would be totally wrong. Instead, they were multiple-velocity and multimode processes, taking place simultaneously on the large scale of nations[2] and on the smaller scale of regions, communities,[3] and institutions. They were affected by multiple variables, such as the type of internet connections available and the technical-political dynamics behind it;[4] also at play could have been localisms, which in scientific environments could grow, for example, around outstanding projects or research centers until the broader diffusion of the internet in the 1990s – and beyond. Such a techno-cultural complexity evokes Hyman's lucid definition of "virtual tribalism […] rather than a global village"[5] and, for as much as the growing internet – especially, the early internet – is concerned, seems to challenge the famous concept McLuhan had forged for the interconnected world.

---

[1] E.g.: Janet Abbate, Inventing the Internet (MIT Press, 1999), 208-9, 181-88; Kevin Driscoll, *The Modem World. A Prehistory of Social Media* (Yale University Press, 2022), 5-11; Kevin Driscoll and Camille Paloque-Bergès, "Searching for Missing 'Net Histories,'" *Internet Histories* 1, no. 1-2 (2017), https://doi.org/10.1080/24701475.2017.1307541.

[2] Ronda Hauben reports that "Dik Winter, from Amsterdam […], describes how the first cross Atlantic Usenet link was delayed until 1982/83 because of the difficulty of acquiring an auto dialer modem that conformed to European standards. 'In Europe,' he writes, 'the two people responsible for the link were […] at the Mathematisch Centrum, a research site in Amsterdam […]". On the other side, "Hagen writes that European Unix users who met in European DEC meetings began to do networking in the late 1970's," initially working from centers situated in the UK, Denmark and the Netherlands (Ronda Hauben, "On the Early Days of Usenet: The Roots of the Cooperative Online Culture", in *Netizens: On the History and Impact of Usenet and the Internet*. IEEE Computer Society Press, 1997, chap. 10, https://www.columbia.edu/~rh120/ch106.x10, accessed January 10, 2024); see also Peter Kirstein, "Early Experiences With the Arpanet and Internet in the United Kingdom," *IEEE Annals for the History of Computing* 21, no. 1 (1999): 38-42.

[3] For Usenet newsgroups and communities see e.g. Nancy K. Baym, "From Practice to Culture on Usenet," *The Sociological Review* 42, no. 1 (1994): 30 and throughout the article. For BBSs see Driscoll, *The Modem World*, 132-69.

[4] Michael Hauben, "The Social Forces Behind the Development of Usenet," in *Netizens,* archived August 4, 2016, at https://web.archive.org/web/20160804110914/http://www.columbia.edu/~hauben/book/ch106.x03; Janet Abbate, *Inventing the Internet* (MIT Press, 1999), 186-87 and throughout the book; Peter Kirstein, "Early Experiences", 40-41 and beyond.

In the astrophysics environment it is notable that as late as 1996 the 130-page guide "Yellow NetPages™ – USENET Newsgroups," which listed about 15000 newsgroups worldwide,[6] included dozens of astronomical newsgroups of national or institutional scope, when the two international newsgroups under examination were born years before.[7] Also, in 1994 Andernach and colleagues informed that "the 'ADASS' news hierarchy (for Astronomical Data Analysis Software and Systems) has been established in parallel to the Usenet news. This set of newsgroups is intended as a forum for discussion of astronomical data analysis software. These newsgroups were established by the IRAF Group at NOAO. […] The hierarchy currently covers the ADASS Conferences and discussions about IRAF-related software, but other groups are encouraged to establish additional subgroups."[8]

All the same, the creation of sci.astro and sci.astro.research seems to mark a significant milestone in the virtualization of the astrophysical communication due to the reach of an international scale and to their many-to-many working mode,[9] which calls forth the idea of a wide-scale learning community, and a very early one. Actually, for as much as it results and we'll see in some more detail below, it also seems to represent one of the first cases of wide-area Usenet newsgroups uptaken by scholarly communities worldwide.

---

[5] Hyman, "Twenty Years," 19.

[6] "Yellow NetPages - Usenet Newsgroups", Aldea Communications, copyright 1996, archived June 21, 2022, at https://web.archive.org/web/20220621103308/https://www.math.utah.edu/phonebooks/yellnews.pdf.

[7] In 2000 it was maintained that the astrophysics newsgroups listed in the disciplinary database "AstroWeb" could be quantified in thirty-one (Daniel Egret, Robert J. Hanisch, and Fionn Murtagh, "Search and Discovery Tools for Astronomical On-Line Resources and Services," *Astronomy and Astrophysics Supplement Series* 143, no. 1 (2000): 139, https://aas.aanda.org/articles/aas/pdf/2000/07/ds1834.pdf).

[8] Heinz Andernach, Robert J. Hanisch, and Fionn Murtagh, "Network Resources for Astronomers," *Publications of the Astronomical Society of the Pacific (PASP)* 106 (1994): 1192, https://iopscience.iop.org/article/10.1086/133497/pdf.

[9] Notably different from that of, for example, the earlier one-to-one email.

## 4. *Research method and sources*

Web historian Camille Paloque-Bergès conveniently recalls the teaching of the "Annales" School when bolstering the need to "integrat*e* […] a growing variety of sources" to gain well-grounded research in web history.[1] This approach looks mandatory, considering how deficiently and fortuitously networked messages have often been preserved before becoming research objects: Valérie Schafer talks about "black boxes, which seldom allow to measure the loss of data and the representativeness of the preserved elements."[2] These observations are specially appropriate for newsgroups: "the main challenge regarding Usenet archives for historians and social scientists is their accessibility, fragmentation and non-exhaustivity: not only are there holes in the archives […] but there are also several collections with concurrent data [… ].”[3]

[1] "Les archives de courrier électronique sont considérées comme 'un problème très contemporain de préservation de la mémoire' (Bergeron *et al.*, 2014, p. 212). L'intégration d'une variété croissante de sources [….] est un fait accepté de l'historiographie depuis au moins l'école des Annales." ["The archives of emails are considered as 'a very contemporary issue of preservation of memory' (Bergeron *et al.*, 2014, p. 212). The integration of a growing variety of sources [….] is a fact accepted by historiography at least since the Annales school" (Camille Paloque-Bergès, *Qu'est-Ce Qu'un Forum Internet? Une Généalogie Historique Au Prisme Des Cultures Savantes Numériques* (OpenEdition Press, 2018), 88, https://doi.org/10.4000/books.oep.1843, English translation by the author). See also Driscoll and Paloque-Bergès, "Searching for Missing".

[2] "L'historien […] doit aussi composer avec des boites noires, qui permettent rarement de mesurer la perte de données et la réprésentativité des éléments sauvegardés." ["the historian […] has also to deal with black boxes, which seldom allow to measure the loss of data and the representativeness of the preserved elements."] (Valérie Schafer, "Les Réseaux Sociaux Numériques d'avant….," *Le Temps Des Médias* 2, no. 31 (2018): 134, https://doi.org/10.3917/tdm.031.0121, English translation by the author. In a similar vein, Smith warns that "since every local news server's feed is partial (almost no single news server takes every newsgroup), a truly complete picture of the Usenet may be impossible to generate." (Marc A. Smith, "Invisible Crowds in Cyberspace: Mapping the Social Structure of the Usenet," in *Communities in Cyberspace*, ed. by Marc A. Smith and Peter Kollock (Routledge, 1999), https://courses.ischool.berkeley.edu/i290-12/f06/smith_invisible_crowds.pdf, 10).

In full adhesion to these views and to the consequent recommendations,[4] the present research combines multiple sources: 1. human sources, constituted both by (1.a.) the contribution of four individual astrophysicists, two of which had a role in the maintaining of one or the other resource; and by (1.b.) the results of a survey conducted among Italian astrophysicists and technologists in May 2023, dedicated to their experiences as users of CMC tools and newsgroups; 2. preserved online archives of Usenet newsgroups.

In fact, also the literature has been tentatively explored, but it has not resulted to be particularly fruitful.

## 4.1. *Sources: previous literature on the subject*

To our knowledge, no comprehensive studies document the process of creation and uptake of CMC tools within the astrophysics community, not even regarding the major messaging systems. The good overview by Harley et al., based upon interviews with astrophysicists, sheds light on this community’s uptake and leanings as for some online communication services, but, generally, not in a historical perspective proper and anyway not about newsgroups.[5] In con-

[3] Paloque-Bergès, “Usenet as a web archive,” 248. See also Driscoll and Paloque-Bergès, “Searching for Missing;” Valérie Schafer, “Les Réseaux Sociaux Numériques,” 133-34.

[4] “Crucially, one should never trust one source or one archive, but should confront it with other documents, keeping in mind what each source […] leaves out […]”. Driscoll and Paloque-Bergès, “Searching for Missing”.

[5] Diane Harley et al., “Astrophysics Case Study,” in *Assessing the Future Landscape of Scholarly Communication: An Exploration of Faculty Values and Needs in Seven Disciplines*, by Diane Harley et al. (UC Berkeley, Center for Studies in Higher Education, 2010), http://escholarship.org/uc/item/15x7385g.

trast, high-energy physics seems to have received more attention.[6]

In fact, the astrophysical community produced a sizeable number of publications to spread the knowledge of specific internet resources inside in the early 1990s, after the diffusion of the World Wide Web.[7] The nature of this literature, though, can generally be identified as practical information, typically aimed at optimizing the retrieval and management of astrophysical data through online databases and specialistic websites. Apparently, mere communication outlets – such as newsgroups – were left a negligible role, which is explained by Barjak with the assumption that "astronomers […] rely mostly on impersonal electronic information sources ([…] archives, databases, […]). This probably reflects the fact that in astronomy (and physics) in particular, large-scale databases and archives were established ar a very early stage".[8] Often, this litera-ture has to be found in grey literature venues (newsletters, technical bulletins) or retrieved in conference proceedings dedicated to different subjects. For all these reasons, the short account that follows is tentative.

---

[6] E.g.: Paul Ginsparg, "First Steps Towards Electronic Research Communication," *Computers in Physics* 8, no. 4 (1994): 390-96; Bruce V. Lewenstein, "Do Public Electronic Bullettin Boards Help Create Scientific Knowledge? The Cold Fusion Case," *Science, Technology, & Human Values* 20, no. 2 (1995): 123-49; about physics at large, Uwe Matzat, "Academic Communication and Internet Discussion Groups: Transfer of Information or Creation of Social Contacts?," *Social Networks* 26, no. 3 (2004): 221-55.

[7] E.g.: André Heck and Fionn Murtagh, eds., *Intelligent Information Retrieval: The Case of Astronomy and Related Space Sciences* (Springer, 1993); Heinz Andernach, "On–line Data and Information in Astronomy or Where to Find Astronomical Information without Scanning the Bookshelves of the Library," *IAC Technical Note*, no. 1 (1993): 18, https://articles.adsabs.harvard.edu/pdf/1993IACTN…….1A; Heinz Andernach, Robert J. Hanisch, and Fionn Murtagh, "Network Resources for Astronomers," https://doi.org/10.1086/133497; Fionn Murtagh, "Computer Networking in Astronomy," in *Information & On-Line Data in Astronomy,* ed. by Daniel Egret and Miguel Albrecht (Springer, 1995), 235-41; Egret, Hanisch, and Murtagh, "Search and Discovery Tools", 137-43. This happened also by means of dedicated conferences, such as the ADASS conferences abovementioned, born in 1991 ("Past ADASS Conferences", ADASS, archived November 26, 2024, at https://web.archive.org/web/20241126145553/https://adass.org/pastven.html).

[8] Barjak, Franz, "The Role of the Internet in Informal Scholarly Communication," *Journal of the American Society for Information Science and Technology* 57, no. 10 (2006): 1362.

In 1992, astrophysicist Robert Hanisch mentioned sci.astro as one of the five Usenet newsgroups astronomers would find useful, together with four more specialized newsgroups which had been created between 1991 and 1992 ("also of interest to astronomers are several newsgroups in the Usenet news, such as sci.astro, sci.astro.fits,[9] sci.astro.hubble, alt.sci.astro.aips, and alt.sci.astro.figaro.").[10] This was confirmed by Andernach in 1993 and then by Andernach, Hanisch and Murtagh in "Publications of the Astronomical Society of the Pacific", a research journal, in 1994.[11] The latter stated that the newsgroups "of obvious interest to astronomers," now quantified as six, also included sci.astro.research,[12] which had been created in the meantime for "discussions of current professional research topics", as we'll see. Sci.astro is defined as a newsgroup dedicated to "*general* astronomy discussion and information" (emphasis added) which "provides considerable discussion *of amateur and popular astronomy questions*"; all the same, it "is read by many professional astronomers and is one possible forum for technical questions."[13]

[9] "The alt.sci.astro.fits newsgroup was formed in May 1991 as a discussion forum for FITS format-related topics. This "alt" newsgroup was replaced by the registered sci.astro.fits newsgroup in February 1992." ("Archive of early sci.astro.fits and FITSBITS email postings. 1991 to 1999", archived January 23, 2025, at https://web.archive.org/web/20250123171535/https://fits.gsfc.nasa.gov/fits_fitsbits_archive.html). One of the astrophysicists we have interviewed individually remembered: "I was active in the FITS community at the time and I'm sure we used sci.astro.fits for discussions." (C4)

[10] Robert J. Hanisch, "Services Available on the Network," *AAS Newsletter*, no. 62 (1992), https://aas.org/publications/aas-newsletter/nl62/epubsup (accessed 26 Nov 2024). Sci.astro.hubble was born in January 1992, followed two months later by alt.sci.astro.aips. Alt.sci.astro.figaro had recently been launched as at April 8, 1992.

[11] Heinz Andernach, "On-line Data and Information in Astronomy", 18; Andernach, Hanisch, and Murtagh, "Network resources for astronomers", 1191. The latter adds that "a number of newsgroups are devoted to astronomical discussions".

[12] Andernach, Hanisch, and Murtagh, "Network resources for astronomers", 1191.

In the nearby domain of physics, physicist Paul Ginsparg provides a clear though synthetic account of how high-energy theoretical physicists took up CMC tools. He reports that "this community, by the mid-1980s, had already begun highly informal mechanisms of regular electronic information exchange, in turn enabled by concurrent advances in computer software and hardware. […] By the end of the 1980s, virtually all researchers […] were plugged into one or another of the interconnected worldwide networks and were using e-mail on a daily basis."[14] As we'll see, this periodization seems to fit astrophysics as well in the United States, and arguably in the few other Western countries which had been equipped with networked connections early, such as the UK.[15] Nevertheless, even some of the latter seem to have experienced slightly different paces, such as Italy.[16] One of the individually interviewed astrophysicists testifies that in Germany, at the beginning of the 1990s, "email existed, but wasn't used much. One had to log in to the central computer of the university, a machine which one used mainly for running calculations. Thus, most people used it rarely and weren't very familiar with it. The local machines at the observatory didn't have email and, when I first started, weren't even connected to the internet. So, in summary, one could say that in 1992 email (at least in Hamburg) was used, but occasionally, and was not the main method of communication, but the first 'online' one".[17] About newsgroups, our survey within the Italian National Institute for Astrophysics reveals that the majority of those who used them started doing so after 1995 (51.92%), while 28.84% started between 1991 and 1995. Only one person started between 1980 and 1985.

[13] Ibid.

[14] Ginsparg, "First Steps," 390-91.

[15] Kirstein, "Early Experiences", 38. Cosmologist Peter Coles, who started his activity as "a research student" at the University of Sussex in 1985, reports that he "was regularly using email in 1985." The connection was throgh DECnet and "the STARLINK system in use at Sussex and throughout the UK was […] VAX-based." ("Personal internet history", (blogpost) Peter Coles, https://telescoper.blog/2023/04/09/personal-internet-history, accessed November 26, 2024).

[16] The sixty-seven = > 46 years old respondents to the questionnaire administered to the researchers of the Italian National Institute for Astrophysics (source 1.b. above) indicate that only 16.42% started using emails between 1980 and 1985. As Kirstein put it, "[…] <for> the Germans, Italians, and Norwegians […] in the late 70s, *the* growth of national research networks was much slower [than for the Britons] […] For this reason, it was not possible for a significant academic involvement from those countries with their US colleagues, until USENET, EARN, and other similar internet developments took off in the middle 1980s." (Kirstein, "Early Experiences", 42-43).

Again about physics, based on his inquiry made in 1998-99, Matzat maintains that researchers in physics and in mathematics are among the few scholars using newsgroups.[18] Still, this author's sample is geographically limited to Dutch and British universities.

Interestingly, Lewenstein explores the use high-energy physicists made of a specialized Usenet newsgroup between 1989 and 1992 during the "Cold fusion saga."[19] Due to their relevance, his main points will be discussed in the final section.

### 4.2. *Sources: astrophysicists*

As anthropologist Sharon Traweek put it, "key informants are crucial; they are people with whom one can tryout tentative interpretations and hypotheses. People who are interested in con-sciously reflecting on their own culture tend to be atypical within it, whether leaders, geniuses […]; they are willing to reflect on the differences between themselves and their fellows."[20] These informants are even more precious in a hard-science environment,[21] where getting support for studies in the area of social sciences and humanities deserves special thanks.

[17] C2.

[18] "The use of newsgroups is almost exclusively restricted to researchers within mathematics (10% of all mathematicians), mechanical engineering (9%) and physics (3%)" (Uwe Matzat, "Academic Communication," 236).

[19] Bruce V. Lewenstein, "Do Public Electronic Bulletin Boards Help Create Scientific Knowledge?: The Cold Fusion Case," *Science, Technology & Human Values* 20, no. 2 (1995): 123-49.

As it was discovered, the founder of the two newsgroups under examination passed away in January 2022, before contacting him was attempted. Unfortunately, also the researcher who "rebuilt" the specialistic newsgroup in 1994 couldn't be contacted. Luckily, though, support was received from four astrophysicists who agreed to be interviewed individually.[22] One of them has been a co-moderator of sci.astro.research for eight years and another one has been the FAQ maintainer for sci.astro about from 1995 to 2000;[23] all of them contributed to the two newsgroups. To these, we can add the answers of 67 anonymous astrophysicists and technologists in the age tier = > 46 years (1.b.), who conveyed their experiences through a dedicated online survey.[24] The fact that sci.astro and sci.astro. research's predecessors were born about forty years ago, as we'll see, poses anyway significant limitations to the possibility human informants have to contribute to their history. A common remark from the individual interviewees was the difficulty of remembering accurately after many years.[25] Similar difficulties emerged from the respondents to the online survey, specifically about the names of the newsgroups they had used (96% of respondents do not remember them), the duration of their experience (62.50%) and, to a lesser extent, the reason why they had started using newsgroups[26] – although they seem to keep a clear global appraisal of their experience. Under these circumstances, online archives supplement the human contribution invaluably.

[20] Sharon Traweek, *Beamtimes and Lifetimes*, 13.

[21] Traweek lucidly detected "the tremendous force of the division […] between outsiders, no matter how well-informed, and insiders" (*Beamtimes and Lifetimes,* 14).

[22] C1, C2, C3, C4. Three of them were interviewed by email, one by a telephone call.

[23] "When did I get involved [in the management of sci.astro]? In the early 1990s. I am not sure of the exact year, and there may not have been a specific 'start date.' It may have been more of an evolution, in which I participated, but then became more and more active over time." (C1).

[24] 67 people, working at the Italian National Institute or Astrophysics. 95.52% are Italian; 1.49% from other European countries; 2.99% from outside Europe.

## 4.3. *Sources: the digital archives*

The online archives that prove to be relevant for the present research are the different existing online collections of our two newsgroups and those of their predecessors – which will be better explained later on –, as well as the online archives of what will be called *meta-newsgroups*. I define the latter as trans-disciplinary newsgroups spreading information on newsgroups' major organizational aspects (new entries, moderators' contact addresses or, e.g., policies). The importance of meta-newsgroups is paramount, as they supplement the information we can get about individual newsgroups under many foundational respects.

According to Ronda Hauben, the practice of archiving posts from Usenet newsgroups started on the basis of individual interests "by the early 1990s."[27] While warning that "not only the very earliest Usenet posts, before […] 1981 […] but even some of the posts in the 1980s are still lost,"[28] Mieszkowski correctly brings this prac-tice much forward, as she mentions that Canadian computer programmer "[Henry] Spencer *started archiving in 1981.*"[29]

[25] "Since this is over 30 years ago, I don't recall much about sci.astro" (C4); "my memory might not be entirely accurate" (C1); the same caveat was made in the beginning of the telephone interview (C3).

[26] 12.02% of respondents.

[27] When "individual Usenet participants archived the posts of some Usenet newsgroups."(Ronda Hauben, "Commodifying Usenet," 64).

[28] Katharine Mieszkowski, "The Geeks Who Saved Usenet".

Importantly, R. Hauben warns that newsgroups' "posts [...] circulated until their expiration date. Each site could set its own date for the expiration of the posts, but they all expired. Consequently, a user would contribute a post and it would be sent out across the globe, but it would expire and disappear *from each node* on the network *on different* but set *dates*."[30] Thus, problems arise not only with finding suitable archives for the topics of interest, but – as a specific issue for newsgroups – also with the actual coverage of the archives retrieved, independently of the apparent one. For the same reason, the coverage can be different in archives apparently covering the same years. Substantially, due to what some web historians perceive as "a [...] sort of anti-memory design,"[31] all the archives of Usenet newsgroups which are available may suffer from non-negligible and random gaps, different from one another, which make it challenging to build on them unless they are carefully analyzed, compared to one another[32], and combined.

In order to find suitable archives of Usenet newsgroups, we have initially built on Baumann,[33] who reports the existence of

---

[29] Mieszkowski, "The Geeks", emphasis added; see also Paloque-Bergès, "Usenet as a web archive," 236-37. About Spencer, the "legendary Unix hacker" who by 1981 "ran the computer facility at the University of Toronto's zoology department", he started saving newsgroups posts and eventually obtained an archive of "141 tapes, most of which held 120 megabytes of posts" (Mieszkowski, "The Geeks").

[30] Ronda Hauben, "Commodifying Usenet," 64 (emphasis added). See also Driscoll and Paloque Bergès, "Searching for Missing".

[31] Driscoll and Paloque Bergès, "Searching for Missing".

[32] "Crucially, one should never trust one source or one archive, but should confront it with other documents, keeping in mind what each source [...] leaves out [...]" (Driscoll and Paloque-Bergès, "Searching for Missing").

[33] Ryan Baumann, "Early Usenet History and Archiving" (blogpost), archived January 29, 2024, at https://web.archive.org/web/20240129120207/https://ryanfb.xyz/etc/2015/02/23/early_usenet_history_and_archiving.html. This text was "originally published: 2015-02-23" and "last modified: 2015-03-11".

five online archives (a): a.1.) Google Groups;[34] a.2.) The Internet Archive's *Usenet Archive of UTZOO Tapes*;[35] a.3.) The Internet Archive's *Usenet Historical Collection*;[36] a.4.) *The Usenet Archive*[37] and a.5.*) A-News Archive* – "Early Usenet News Articles: 1981 to 1982."[38] Very briefly, a.2.) and a.4.) have become unavailable, while the coverage of a.3. results to be too limited for our purposes,[39] so it has been used to a minimal extent. Further online archives have been retrieved independently (b). The most useful

---

[34] https://groups.google.com/ (from the homepage, you can access only by signing in with a Google account). According to Paloque-Bergès, "Google a entrepris l'archivage des groupes de Usenet en 2001, une initiative de conservation de discussions publiques par une entreprise privée […] controversée […] cet effort est aujourd'hui considéré comme une ruine numérique." ("Google has undertaken the archiving of Usenet newsgroups in 2001, a […] controversial […] initiative of public discussion by a private company […] this effort today is considered as a digital ruin.") (*Qu'est-ce qu'un forum internet?,* 92; English translation by the author).

[35] "The UTZOO Wiseman Usenet Archive", David Wiseman, https://archive.org/details/utzoo-wiseman-usenet-archive. "In 2020 after sustained legal demands requesting a set of messages within the Usenet Archive be redacted, and to avoid further costs and accusations of manipulation should those demands be met, the archive has been removed from this URL and is not currently accessible to the public." (ibid.).

[36] "Usenet Historical Collection", created January 21, 2014, http://archive.org/details/usenethistorical.

[37] http://www.theusenetarchive.com/.

[38] "Usenet Oldnews Archive Newsgroups List", https://web.archive.org/web/20000303203929/http:/communication.ucsd.edu/A-News/index.html. At the bottom of the homepage it is stated that "the Usenet Oldnews Archive is Copyright © 1981, 1996, by: Bruce Jones […], Henry Spencer […] <and> David Wiseman".

[39] We found that net.astro is available from 1984 to 1985 (https://archive.org/download/usenet-net/net.astro.mbox.zip) and net.astro.expert for the same years but only for some months (https://archive.org/download/usenet-net/net.astro.expert.mbox.zip).

In the subdirectory https://archive.org/download/usenet-sci, ("files for usenet-sci") we have found sci.astro.research from 1994 to 2013 (on sci.astro.research.mbox.zip) and sci.astro from 2003 to 2013 (on https://archive.org/download/usenet-sci/sci.astro.mbox.zip).

ones have proved to be (b.1.) Josef (Joe) Jarosciak's *Usenet archives*[40] and (b.2.) the apparently untitled resource available at https://altavista.superglobalmegacorp.com/altavista/.[41] Most of these archives result to have built on the same archetypal source, just the unanimously praised pioneering archive created by Henry Spencer in 1981, which covered the crucial, almost initial decade 1981-91: a.1.)., a.2.), a.5.),[42] b.1.)[43] and b.2.)[44] (this common origin doesn't prevent them from retrieving collections of the same Usenet newsgroups which are slightly different from each other).

---

[40] "Usenet archives", archived June 28, 2022, at https://web.archive.org/web/20220628134008/https://www.usenetarchives.com/. It was created in "2020-2021" allegedly "as a way to host groups in a way that'd be independent of Google Groups" (Samantha Cole, "2.1 Million of the Oldest Internet Posts Are Now Online for Anyone to Read," *Vice*, October 13, 2020, archived August 10, 2022, at https://web.archive.org/web/20220810144946/https://www.vice.com/en/article/pky7km/usenet-archive-utzoo-online). See also: Motherboard, "2.1 Million of the Oldest Internet Posts Are Now Online for Anyone to Read," ACM News, October 16, 2020, archived January 29, 2024, at https://web.archive.org/web/20240129175023/https://cacm.acm.org/news/248041-21-million-of-the-oldest-internet-posts-are-now-online-for-anyone-to-read/fulltext#comments; Jozef Jarosciak, "Converting UTZOO Usenet archive from magnetic tapes to MySQL database using Java", dated February 17, 2019, archived January 29, 2024, at https://web.archive.org/web/20240129175516/https://www.joe0.com/2019/02/17/converting-utzoo-usenet-archive-from-tgz-to-mysql-database-java-code/.

[41] As archived on the Internet Archive (November 26, 2024): https://web.archive.org/web/20241126151853/https://altavista.superglobalmegacorp.com/altavista/. The nickname associated is "Neozeed". This resource seems to have been created in October 2021, according to the dates of the files uploaded, as they appear in the main directory.

[42] Mieszkowski, "The Geeks"; David Wiseman recalls this process at https://ia801903.us.archive.org/9/items/utzoo-wiseman-usenet-archive/introduction.txt (accessed November 26, 2024). These files, anyway, seem to have (partly?) migrated also to different online archives (Mieszkowski,"The Geeks"; Driscoll and Paloque-Bergès, "Searching for missing"), through specific paths that are outside the scope of the present paper. So it happens also for a.2.) and a.5.).

[43] Jarosciak, "Converting UTZOO Usenet archive."

[44] As clearly stated on the homepage: "search UTZOO archive", i.e. the archive created by Henry Spencer.

These resources do not represent a comprehensive census of online archives of newsgroups, which would require a thoughtful dedicated study and would most probably result to be utopical. Nevertheless, they seem to provide sufficient information for our research purpose, so they have deemed to be extended enough.

Issues with these resources have been experienced with critical features such as persistence,[45] description of the content,[46] the possibility to search by date, period,[47] author or subject, completeness of the message headers and availability of the threaded conversations. We will not go into further detail about these aspects, whose relevance has been noted by web historians. To summarize, the online archives upon which the present research was built have been: a.1.) (Google Groups); a.3.) (Usenet Historical Collection); b.1.) (Usenet archives);[48] and b.2.). Some useful information also came from a.5.) (A-News archive).

[45] For this reason I have archived as many as possible of the available ones on the Internet Archive (https://archive.org/) and I'm citing them with the url they got as a consequence of this process. The same applies to some other relevant online resources cited in this paper.

[46] "Les archives de Usenet […] sans accompagnement patrimonial explicite, […] sont de fait peu connues et difficiles d'accès et d'exploitation." ["The archives of Usenet […] without explicit description of the content, are in fact little known and difficult to access and use"] (Paloque-Berges, *Qu'est-Ce Qu'un Forum Internet?,* 92, note 7; English translation by the author).

[47] Paloque-Bergès, "Usenet as a Web Archive," 239.

[48] Usenet Archives' interface lets users query by year and by word of text easily, quantifies posts by year, letting you get to them easily but doesn't seem to include the author search and, especially, its content appears to be quantitatively more limited than that, e.g., on GG.

## 5. *In the beginning: net.astro and net.astro.expert (November, 1983)*

At the origins of the international Usenet newsgroups dedicated to astrophysics we find net.astro and net.astro.expert.

The creation of both newsgroups was proposed by US astrophysicist William Lawrence Sebok (July 10, 1951 - January 4, 2022), by that time thirty-two years old and working as a research associate at Princeton University.[1] He did so through a newsgroup post on November 14, 1983,[2] substantially following a general procedure that had certainly been prescribed on February 3, 1982.[3]

---

[1] "A brilliant student, Bill graduated from Tallmadge High School and the University of Akron. He attended graduate school at Caltech, where he earned a Ph.D. in astronomy. Bill was a postdoctoral researcher at Princeton University and spent most of his career in the Astronomy Department at the University of Maryland, College Park, where he was a computer system manager and software expert." ("William Lawrence Sebok. Obituary," Tributearchive.com, archived January 3, 2023, at https://web.archive.org/web/20230103182538/https://www.tributearchive.com/obituaries/23588343/william-l-sebok/ellicott-city/maryland/harry-h-witzkes-family-funeral-home). Further information about Bill Sebok can be found on his personal webpage at the University of Maryland (last modified February 12, 2016), archived March 19, 2016, at https://web.archive.org/web/20160319191855/http://furo.astro.umd.edu/; see also "William Lawrence Sebok", Wikipedia, archived December 2, 2024, at https://web.archive.org/web/20241202130908/https://it.wikipedia.org/wiki/William_Lawrence_Sebok.

[2] The post, with subject "Net.astro", was sent to the meta-newsgroup net.news.group, and to the newsgroup net.space. Respectively: https://web.archive.org/web/20230404122849/https://utzoo.superglobalmegacorp.com/usenet/news008f1/b19/net/space/1643.txt and https://web.archive.org/web/20231017130607/https://utzoo.superglobalmegacorp.com/usenet/news008f1/b19/net/news/group/835.txt. GG preserves one, favourable comment attached to the post on net.space (https://groups.google.com/g/net.space/c/76iybyGt5uQ/m/ejIczvX97XcJ) and six, also favourable, to that on net.news.group (https://web.archive.org/web/20240130134653/https://groups.google.com/g/net.news.group/c/76iybyGt5uQ/m/ejIczvX97XcJ).

[3] "A new newsgroup may be created by simply posting material to the net under a new newsgroup name. However, THIS IS NOT RECOMMENDED! There are limits to the number of newsgroups that can be supported by the net. If you wish to send material to the net, first try to find an established newsgroup

The important A-News Archive, which covers a very early period of newsgroups' existence (May 1981 - May 1982), does not contain any astronomical or astrophysical newsgroup, nor do so some further, early lists of newsgroups I have retrieved, covering the years 1982-83. Thus, it appears to be very likely that net.astro and net.astro.expert were in fact pioneering wide-area initiatives for astrophysics. Not only: although it would be risky to judge the scholarly nature of the newsgroups included in the early lists above before a complete scrutiny of their content, a list of the 144 net.* newsgroups available as at October 1, 1983 we have found[4] seems to show that the groups created before net.astro and net.astro.expert were seldom of a scholarly nature – although some of them have conveyed highly specialistic technical content such as IT issues. We can very roughly estimate that scholarly newsgroups accounted for less than 10% of the total; among them we find a newsgroup for mathematics, net.math, born in February 1982, and another one for physics – net.physics – which was created later that year. This emphasizes the importance of net.astro and, especially, net.astro.expert for the beginning of wide-area scholarly communication at large. As we learn from Smith, more than fifteen years later the newsgroups categorized under the "science" label were 203 worldwide, out of 14347 "carried in the UCLA news server's feed", a subset of the "more than 79000" numbered by "Netscan studies".[5]

that deals with a subject related to that material. If there is no appropriate newsgroup, suggest the creation of a new group via net.news.group. Usually, there will be enough feedback to establish whether there is an audience for the subject that you would like to discuss." https://web.archive.org/web/20231025112028/https://groups.google.com/g/net.news.group/c/IdR9WeVdGc4. The author was Curt Stephens.

[4] "List of Active Newsgroups", Adam Buchsbaum, October 1, 1983, archived January 22, 2024, at https://web.archive.org/web/20240122153634/https://groups.google.com/g/net.news.group/c/ENPFOemh-3g. The list includes also 16 fa.* "groups that are gatewayed to USENET from the ARPANET".

[5] Anyway, among the newsgroups included in the "Netscan studies […] many may be only locally distributed" (Marc A. Smith, "Invisible Crowds in Cyberspace", 10).

In Sebok's words:

> *"I would like to propose the establishment of net.astro. This group would be for topics in and relating to astronomy. It would NOT be about the space program, which is the territory of net.space.*[6] [...] *There is much excitement going on in astronomy and many people with access to the net who could contribute information on what is currently happening (indeed many of those people are making it happen). Many (perhaps, judging from the people at Princeton, I could even say most) of these people keep silent because they are not very interested in the contents of net.space (and often, not very interested in the contents of the rest of the net itself). I am proposing a newsgroup for these people, to bring them out of the woodwork. And I think that news of what is happening in astronomy is exciting enough to be of interest to the general public.* [...] *Perhaps what I am really proposing is a net.astro.wizards, in analogy with net.unix-wizards. Just plain net.astro would then be for questions of the order of "Why does the moon look larger at the horizon?" which would be unwelcome in net.astro.wizards. If amateur astronomers wished to establish a group to discuss topics of interest to them they could call it something like net.astro.amateur. Comments? Please feel free to mail comments to me or post them to this group on the net."*[7]

Sebok was substantially proposing to create two different-level newsgroups, one for tendentially non-specialistic discussion, net.astro, and another one for advanced topics – which he at first hypothesized to call "net.astro.wizards" but later on named "net.

[6] As we learn, net.space already existed by April 19, 1982, was dedicated to "Space programs and research" and was undigested from fa.space.

[7] "Net.astro", William Sebok, November 14, 1983, archived October 17, 2023, at https://web.archive.org/web/20231017130607/https://utzoo.superglobalmegacorp.com/usenet/news008f1/b19/net/news/group/835.txt. Sci.astro was considered to be "useful for public outreach" by one of the astrophysicists we have interviewed individually, at the beginning of his involvement in its management in the early '90s (C1).

astro.expert".[8] He supposed that astronomy amateurs might wish to create a group of their own ("net.astro.amateurs"), being apparently unsure about that.

In an email reply to Sebok about his proposal, US IT expert Douglas Tody informed that he had "talked about setting up an electronic network among the astronomy centers in the past but never got anywhere. Your idea seems like a good way to do it, since there are already a significant number of sites on the unix network."[9]

The founder had started discussing his idea with his colleagues at Princeton University around October 1983.[10] Notably,

> *"the main concern I have been told about in the month of consultation I did before posting the original article was a fear that the level of discussion in this group would be too trivial to be worth following (and contributing to). Thus the idea of a separate subgroup, tentatively named net.astro.wizards. These people are less computer oriented, and thus need more encouragement, than professionals in systems programming (net.unix-wizards) or artificial intellegence (net.ai). Also, as I previously mentioned, amateur astronomers would then be in good po-*

[8] Sebok's change of mind about the name of the new resource is testified by a message of his dated November 21, 1983, archived April, 4, 2023, at https://web.archive.org/web/20230404121504/https://utzoo.superglobalmegacorp.com/usenet/news007f3/b19/net/news/group/866.txt ("Summary of net.astro responses"; it was sent to both net.news.group and net.space).

[9] The email was reposted by Sebok ibid. About Doug Tody, by that time 31 years old, see Robert Hanisch, "Douglas Tody (1952-2022)," *Bulletin of the AAS* 54, no. 1 (2022): 1-5, https://web.archive.org/web/20240131151737/https://baas.aas.org/pub/2022i034/release/1.

[10] "Net.astro", November 14 and 16, 1983 (threaded messages on GG), archived January 30, 2024, at https://web.archive.org/web/20240130160528/https://groups.google.com/g/net.news.group/c/76iybyGt5uQ; "Net.foo and Net.foo.expert," December 18, 1983, (in a thread on GG), archived April 26, 2023, at https://web.archive.org/web/20230426191214/https://groups.google.com/g/net.astro.expert/c/Bs2IqxIZJr0. E.g.: "This article been posted after consultation with the members of the Department here, as well has some of the astronomers at the Institute for Advanced study." (November 14, 1983).

*sition to establish their own group, tentatively net.astro.amateur. Amateur astronomers and professional astronomers are mostly (but not always) different people. There would then be a well defined place, net. astro, to put beginners questions."*[11]

On the same net.astro.expert, Sebok is very clear about his goals, sheds light on the perceived attitude of his environment and mentions a previous achievement of his, the server Astrovax:

*"My goals in the establishment of net.astro were 1) to get astronomers involved in the net 2) to increase the support of astronomers and astronomy sites for net as a whole and to encourage new astronomy sites to join. I know how tenous the initial support for the net really is; if it wasn't for my own action astrovax would not be on the net."*[12]

Apart from this post, traces of these origins and intentions can hardly be found in the first few years of the best-known online archives of newsgroups, GG, whose collection of net.astro spans from November 29, 1983 to October 19, 1986[13] (starts a bit later in different archives; ends the same day on Usenet Archives), while net.astro.expert is there from November 29, 1983 to October 11, 1986.[14] For net.astro, the oldest message preserved on GG[15] is a reply to a not included message, so even this comprehensive collec-tion does not provide us either with the first post, or with the new newsgroup's original scopes, motivations or policies.

[11] "Net.astro", November 16, 1983, https://web.archive.org/web/20240130160528/https://groups.google.com/g/net.news.group/c/76iybyGt5uQ.

[12] "Net.foo and Net.foo.expert", December 18, 1983, https://web.archive.org/web/20230426191214/https://groups.google.com/g/net.astro.expert/c/Bs2IqxIZJr0.

[13] "Net.astro", archived January 10, 2024, at https://web.archive.org/web/20240110081511/https://groups.google.com/g/net.astro.

[14] "Net.astro.expert", archived January 10, 2024, at https://web.archive.org/web/20240110081421/https://groups.google.com/g/net.astro.expert.

[15] "STS-9 launch and Pravda announcement", William L. Sebok, November 29, 1983, archived January 8, 2024, at https://web.archive.org/web/20240108164558/https://groups.google.com/g/net.astro/c/ElWFzupRa6c.

The actual creation of net.astro and net.astro.expert happened through two control messages sent by Sebok, which have been retrieved.[16] Both of them were posted on November 26, 1983. They include the two newsgroups' scopes: "net.astro is for *discussion* of topics in and related to astronomy"; "net.astro.expert is for *informed discussion* of topics in and related to astronomy" (emphasis added). The meta-newsgroup net.news.group confirms that Net. astro wasn't active as of November 15, 1983, but was on December 1.

For net.astro.expert, both Usenet Archives and Google Groups provide what is likely to be the first post, dated November 28, 1983.[17] Its author, Steve Grandi of Kitt Peak National Observatory, notices that "net.astro.expert has just magically appeared this morning" (to which Sebok replies: "For me it didn't just magically appear but took more than a month of work to bring it into existence. [...]").

Grandi looks eager to experience the new tool:

[16] "Cmsg newgroup net.astro", Bill Sebok, November 26, 1983, archived April 4, 2023, at https://web.archive.org/web/20230404123443/https://utzoo.superglobalmegacorp.com/usenet/news008f1/b19/control/463.txt; "cmsg newgroup net.astro.expert", Bill Sebok, November 26, 1983, archived April 4, 2023, at https://web.archive.org/web/20230404123945/https://utzoo.superglobalmegacorp.com/usenet/news008f1/b19/control/464.txt. They have been retrieved also on Usenet Archives. About control messages and their functions see https://en.wikipedia.org/wiki/Control_message, accessed November 24, 2024.

[17] "Astronomical computing", Steve Grandi. On Usenet Archives: https://web.archive.org/web/20240115163009/https://www.usenetarchives.com/view.php?id=net.astro.expert&mid=PDI3MEBrcG5vLlVVQ1A%2B (thread), archived January 15, 2024; on GG it is available with one additional comment attached, and a different time-zone stamp (https://web.archive.org/web/20231107100852/https://groups.google.com/g/net.astro.expert/c/GLym0fyl-iE, archived November 7, 2023).

> *"Since net.astro.expert has just* […] *appeared* […] *I thought I would try to get an interesting discussion going. Given that the medium of the discussion is a computer network, my topic may seem rather obvious – the state of computing in professional astronomy. Let me pose two questions and make a few comments about each one – Should a national astronomy supercomputing center be set up?* […] *Are astronomy graduate students getting a proper grounding in software engineering?"*[18]

Interestingly, administrator Greg Woods replies that these questions had been asked in the wrong group: in his opinion, socio-technical subjects were not suitable for the expert group. As he put it,

> *"From my understanding of these two groups [net.astro and net.astro.expert], this topic does not belong in the expert subgroup, which was intended for technical discussions *directly* related to astronomy. This is sort of a peripheral topic (albeit an important one!), and so belongs in the more general group net.astro. I have posted this article to both groups in an attempt to move the discussion to where I think it belongs."*[19]

The borders between net.astro and net.astro.expert didn't appear to be completely clear to other readers, too, as we see e.g. in a post Sebok rebuts to on December 18.[20] Following, the rationale on the basis of net.astro.expert is repeated:

[18] https://web.archive.org/web/20231107100852/https://groups.google.com/g/net.astro.expert/c/GLym0fyl-iE.

[19] Greg Woods, November 29, 1983, ibid. This comment is missing in the Usenet Archives' collection.

[20] "Do you read net.astro.expert and not net.astro, or net.astro and not net.astro.expert? As long as few or no people actually read one group and not the other, there's no need for separate groups." (Dave Sherman (quotation), in "Net.foo and net.foo.expert", Bill Sebok, December 18, 1983, archived January 3, 2023, at https://web.archive.org/web/20230103183254/https://groups.google.com/g/net.astro.expert/c/Bs2IqxIZJr0). In the founder's reply, "I think his point that most things were posted to both groups was ill founded anyway. I see one such article in our system now. I don't recall seeing more than two other double postings since the groups were founded."

*"Before I proposed net.astro I talked to the astronomers here and at the Institute for Advanced Study about the idea.* […] *I could see that the main excuse these people might give for not contributing was that the level of the discussion was too low and that it was wasting their time.* […] *I was very afraid that this might be an example of the discussion on net.astro. I also have not been particularly impressed with net.physics. Thus the idea of net.astro.expert to give the experts a place to speak. It has been my hope (and still is) that some of the interesting discussions I hear at lunches here and elsewhere I might begin to hear on the net.* […]. *If net.astro.expert had been created later [than net.astro] the astronomers would have written off the whole idea of the net."*[21]

In fact, what Sebok had in mind for net.astro.expert was not simply a resource for advanced information sharing; instead, ideally, he envisioned a place for debating interesting ideas, a venue for lively professional exchanges.

Three weeks after the debut, Sebok informs that the traffic on net.astro.expert was not very high and anticipates he would take some remedial action about that.[22]

On November 4, 1985, he informed that he had made a manual addition to the newsgroup, putting it in contact with a technical mailing list of US astrophysicists, the VLBI mailing list. Interestingly, Sebok decided that only selected contents would be conveyed from it to net.astro.expert, whereas no selection would be made from net.astro.expert to the technical mailing list:

*"As an experiment I have started to run a manual gateway between net.astro.expert and the VLBI (Very Long Baseline Interferometry)*

[21] Ibid.

[22] "I have not been overwhelmed by the response yet but I very strongly think it is too early to call it a failure. I have some ideas to stimulate things that I might carry out after I'm back from Christmas vacation. […]" (ibid.). He also calls from more contributions from his peers: "Astronomers, please get your collegues to post something (or better yet, post something yourselves). As Dr. John Bahcall says at the Tuesday lunch at the Institute for Advanced Studies, 'Tell us something interesting'" (ibid.).

*mailing list on the NRAO decnet (tc_vlbi%pho…@cit-hamlet.arpa). This will put another group of real astronomers in contact with net. astro.expert. In the net.astro.expert – > VLBI mailing list direction I will pass everything. However the VLBI mailing list often consists of updates to various pieces of astronomical software. Most of the postings of this type I will weed out, although I may let some get through to give the rest of the net some information on the state of professional astronomical software."*[23]

It is difficult to say how popular net.astro.expert became in the astrophysics community. The most comprehensive collection of this newsgroup we have consulted, that of GG, contains 193 posts and comments distributed in four years (Usenet Archives stops at 86). The distribution is as follows: 11 in 1983 (from November 29); 70 in 1984; 85 in 1985; 27 in 1986 (until October 11, for the reasons we'll see in section 6). It seems reasonable to suppose that net.astro.expert was sufficiently popular for that period, but the uncertainty about how many posts and comments were preserved and our scarce knowledge of how much exactly astrophysicists could connect to CMC tools around 1983 suggest caution in judging this aspect.

Certainly, according to the messages preserved on GG, it results that net.astro.expert's activity never went outside of the US environment, which is strongly related to the situation of the networked connections available at that time.[24] A guest post from an Australian researcher, dated July 18, 1985 and forwarded by Steve Grandi, informs that "we don't get (nor can post to) net.astro or net.astro.

[23] "Experimental NRAO decnet VLBI mailing list gateway", Bill Sebok, November 4, 1985, archived December 13, 2022, at https://web.archive.org/web/20221213101454/https://groups.google.com/g/net.astro.expert/c/evzDq-jvB62U.

[24] See above, note 2 on par. 3.

expert anywhere in Australia, given the costs of transpacific phone calls."[25]

[25] "Astronomical software", Steve Grandi, July 19, 1985, archived January 10, 2024, at https://web.archive.org/web/20240110094458/https://groups.google.com/g/net.astro.expert/c/Ya7xz3hCOC4/m/9gVv7Dm5In0J. Schafer rightly recalls that "dans les années 1980 et partie de la décennie 1990 la facturation à la durée est la règle" ["During the 80s and part of the 90s, billing methods based on duration are the rule"] (Schafer, "Les Réseaux", 128; English translation by the author), certainly in France but also in other countries (see also R. Hauben, "On the Early Days of Usenet").

## 6. *The Great Renaming (1986) and the birth of the unified sci.astro*

The Great Renaming was "one of the most important events to take place during the 'Golden Age' of Usenet" and of newsgroups;[1] a turning point, which brought to a change in newsgroups' names, classification and organization.

This process was announced in two crucial posts by one of its protagonists, who signs as "Rick" – in all likelihood, internet pioneer and influential administrator Rick Adams –,[2] on the meta-newsgroup net.news.group, on August 11, 1986. The first post describes the modes and scheduled timing of the Renaming, as they had been decided by the powerful board of "approximately forty" administrators – also known as "the Backbone Cabal"[3] – of which "Rick" was a member.

As we learn, the transition from old to new names would take place "in two phases. Roughly half of them will be created in mid-September [1986], the rest at a later time when we feel the bugs have been worked out [...]. At a later date (probably December) the "backbone" will simply stop carrying the "net" groups."[4]

---

[1] Lee S. Bumgarner, "The Great Renaming FAQ. Part 1," n.d. (between 1994 and 2006), archived July 9, 2008, at https://web.archive.org/web/20080709061430/http://danflood.com/cs-content/cshistory/csh_greatrename1.html. See also Henry Edward Hardy, "The History of the Net. Master's Thesis".

[2] About Adams ("net.god, future UUnet founder and Bill Gates pal") and his role in the Great Renaming, see Bumgarner, "The Great Renaming." See also "The Living Internet" (copyright 2000), archived December 4, 2022, at https://web.archive.org/web/20221204083707/https://www.livinginternet.info/u/ui_modern_renamingfaq.htm; from the copyright page, the author results to be William Stewart (https://web.archive.org/web/20220126081553/https://www.livinginternet.info/tcopyright.htm).

[3] "The Backbone Cabal [...] was made up of a small group of male computer experts in their 20's and 30's" (Bumgarner, "The Great Renaming. Part 1"; see also parts 2-4.)

[4] "Newsgroup renaming scheme (1 of 2)", Rick, August 11, 1986, archived December 6, 2022, at https://web.archive.org/web/20221206155428/https://groups.google.com/g/net.news.group/c/rhZVfEQpyPA.

Following, and importanly to us, Rick Adams lists the newsgroups and flanks their current names with the new ones they would assume after the Renaming process.

As we learn, the Backbone Cabal's decision about net.astro and net.astro.expert was not only to change their names according to the prefix "sci." created for "Science, Research and Technology," but also to merge them in one newsgroup – which seems to have happened only very rarely (thus perhaps meeting the perplexities expressed by some readers about their perceived similar content?). The name of the unified newsgroup would be "sci.astro"[5], which remained unchanged until 1994.

Around 1995, after the split we'll see, the readers of sci.astro were "51000" and sci.astro appeared to be a victim of its success: messages were estimated to be "3170" per month and "106" per day.[6]

---

[5] Ibid.

[6] "Sci.astro", archived December 6, 2022, at https://web.archive.org/web/20010224093004/http://www.ibiblio.org/usenet-i/groups-html/sci.astro.html. This information is provided by the "Newsgroup Info Center", "copyrighted 1995 by Kevin Atkinson" (https://web.archive.org/web/20010413165900/http://www.ibiblio.org/usenet-i/info/copyright.html, archived April 13, 2001).

## 7. *A new worldwide scenario: globalization, specialization and need for moderation in the 1990s*

In the 1990s, a rapid and massive expansion of the number of people being able to connect to the Internet worldwide was experienced.[1] Baym, based on data by Rick Adams, maintains that this tide affected also the Usenet newsgroups, starting in 1989 with a further acceleration in 1992.[2]

This evolution led to a rapid decline of the previously dominating early-internet communication culture and practice, which were strongly affected by the need to prevent the waste of valuable bandwidth.[3] Avoidance of redundancy, avoidance of cross-posting, making sure to read previous messages before posting, for example, were repeated prescriptions in the early period. The more and more frequent violations of these rules led, among other things, to the resignation of a first-magnitude member of the Backbone Cabal, Gene Spafford,[4] from his administrator role.[5] As a consequence of this

[1] E.g.: Abbate, "Inventing the Internet," 181; Helen V. Milner, "The Global Spread of the Internet: The Role of International Diffusion Pressures in Technology Adoption," in *2nd Conference on Interdependence, Diffusion, and Sovereignty* (UCLA, 2003), https://ssrn.com/abstract=2182083, accessed November 24, 2024; James Curran, "Rethinking Internet History," in *Misunderstanding the Internet*, by James Curran, Natalie Fenton, and Des Freedman (Routledge, 2012), 35.

[2] Baym, "From Practice to Culture on Usenet," 31. As documented, this phenomenon affected the number of newsgroups, that of articles and the number of sent bytes.

[3] E.g.: about mailing lists, "it was considered bad etiquette in the early 1990s to engage in a debate, due to precious bandwidth consumption. […] the first actual flame war […] is occurring no sooner than June 1993" (Hocquet and Wieber, "Mailing List Archives," 46).

[4] About Eugene H. Spafford, "professor […] in Computer Science at Purdue University, where he has been a member of the faculty since 1987", see "Narrative Bio for Spaf", copyright 2004-2018 by E. H. Spafford, last modified May 18, 2024, archived May 29, 2024, at https://web.archive.org/web/20240529103449/https://spaf.cerias.purdue.edu/narrate.html.

[5] As he explained on April 29, 1993: "starting several years ago […] I have had a growing sense of futility: people on the net can't possibly find the postings use-

process, a vast amount of poor-quality, misdirected or just spamming-type messages started appearing,[6] especially on those newsgroups which weren't moderated. Sci.astro was one of them: hence, probably, the need to have available a newsgroup where contents would be checked before being disseminated to the readers worldwide.[7] Some of the readers' comments are extremely clear: "too many full-time reseachers are frightened away by the low signal-to-

---

ful, because most of the advice in them is completely ignored. People don't seem to think before posting, they are purposely rude, they blatantly violate copyrights, they crosspost everywhere, use 20 line signature files, and do basically every other thing the postings (and common sense and common courtesy) advise not to. Regularly, there are postings of questions that can be answered by the newusers articles, clearly indicating that they aren't being read. 'Sendsys' bombs and forgeries abound. […] Reason, etiquette, accountability, and compromise are strangers in far too many newsgroups these days." ("Spaf's farewell letter", Eugene Spafford, April 30, 1993 (reposted January 1, 2015), archived December 14, 2022, at https://web.archive.org/web/20221214153737/https://danflood.com/cshistory/spafs-farewell-letter/). As another early internet expert, Lauren Weinstein, wrote on October 25, 1990, "the net has tended only to address the traffic *volume* issues, while as a whole not wanting to worry about the "quality" issues. It's the increase in volume, with if anything a continuing decay of the signal/noise ratio, which has driven many of the "old-timers" (including myself) into much more restricted reading of and participation in the net than some years ago." ("Mail relay", Lauren Weinstein, in a thread; archived December 27, 2022, at https://web.archive.org/web/20221227101821/https://danflood.com/cs-content/cshistory/csh_usenet3.htm). About Weinstein: "Lauren Weinstein (technologist)", Wikimedia Foundation, last modified November 17, 2024, https://en.wikipedia.org/wiki/Lauren_Weinstein_(technologist).

[6] "The apparition of spam from the mid-90s on" is confirmed by Hocquet and Wieber, 41. Thagard recalls how different the situation was between moderated newsgroups, "leaving entries that are likely to be relevant to researcher's work", and unmoderated ones, "often fill up with junk". Significantly, he exemplifies through the difference between sci.physics.research and "unmoderated, junk-laden groups such as sci.physics." (Paul Thagard, "Internet Epistemology: Contributions of New Information Technologies to Scientific Research," © 1997, http://cogsci.uwaterloo.ca/Articles/Pages/Epistemology.html, accessed November 26, 2024).

[7] See Paloque-Bergès, *Qu'est-Ce Qu'un Forum Internet?*, 47.

noise in this group";[8] "sci.astro is getting too filled up with junk to check regularly if I'm busy."[9]

According to Mauldin, in 1991 "by far the bulk of the SCI-ASTRO traffic are discussions of individual questions about astronomy and astrophysics"; "a significant portion of the articles are posted by individuals buying and selling equipment or asking for advice about buying equipment."[10]

## 7.1. *The birth of sci.astro.research (1994)*

Sci.astro.research (https://groups.google.com/g/sci.astro.research on GG) was created in May 1994 as a moderated newsgroup, in addition to sci.astro.

The idea was launched by US physicist Martin E. Sulkanen, who by that time worked at NASA's Marshall Space Flight Center, through a post on sci.astro on February 15, 1994.[11] His aims seem to be aligned with those by William Sebok inasmuch a specific discussion area would be provided for professional astrophysicists. As we'll see, though, things are more nuanced.

In Sulkanen's call to the readers, we read:

---

[8] Comment by a British astrophysicist, February 19, 1994 (in a thread), archived December 27, 2022, at https://web.archive.org/web/20240117141434/https://groups.google.com/g/sci.astro/c/jyxdPTquAjw/m/5x8x4suBzNEJ.

[9] Ibid.

[10] Michael L. Mauldin, "Empirical studies," chap. 6 in *Conceptual Information Retrieval. A Case Study in Adaptive Partial Parsing* (Springer, 1991), 110.

[11] "Is it time for a moderated research subgroup?", Martin Sulkanen, https://web.archive.org/web/20230103180346/https://groups.google.com/g/sci.astro/c/jyxdPTquAjw/m/5x8x4suBzNEJ, followed by a thread of twenty-five comments (they are fifteen on Usenet Archives: https://web.archive.org/web/20221227180904/https://www.usenetarchives.com/view.php?id=sci.astro&mid=PDJqcXY3ZSRoc2xAYXZkbXM4Lm1zZmMubmFzYS5nb3Y%2B, archived December 27, 2022).

*“Dear reader, the newsgroup sci.astro is an unmoderated group for discussion and dissemination of information putatively related to the fields of astronomy, astrophysics, amateur astronomy, astronomy education, and space and physics related topics. The wide popularity of astronomy has made sci.astro a very active newsgroup, with a substantial daily bandwidth of articles on diverse topics.*
*However* […] *the daily bandwidth is quite large, with a correspondingly small “signal-to-noise” ratio of postings that are of relevance to the activities of research astronomers. In addition, sci.astro is also subject to its share of abuse by (cross)posting of articles of minimal or imaginary relevance to astronomy and astrophysics. Thus,* the reaction of many research astronomers to sci.astro has been to avoid it entirely. This has an adverse effect on the timely dissemination of information and related discussion*, for example, on supernovae and other transient phenomena* […]*, as well as providing effective communication between researchers on topics of mutual interest.*
*I would like to suggest the creation of a new moderated newsgroup, sci.astro.research. The manifesto for this group would be similar to that of sci.physics.research* […] *in its moderation philosophy. Postings to the group would be confined to astronomy/astrophysics-related research, but one could imagine that would include questions about hardware, software, and postings of amateur observations of variable stars, comets, and supernovae (something that would be of the quality of an IAU telegram).*
*Also included would be preprint/reprint lists (the NRAO (un)rap sheets), and news summaries in astronomy & physics (“Physics News Update”, etc.).*
*However, I heartily encourage suggestions and opinions on the degree of moderation to sci.astro.research.*
*What would the effect of s.a.r have on s.a? Actually, I think s.a would benefit from the existence of s.a.r… more astronomers would be likely to use news and contribute to both groups.*
*Please send your suggestions and comments to me* […]*. I would like to pursue a formal RFD [Request For Discussion] for sci.astro.research within the next few weeks.”*[12]

[12] “Is it time for a moderated research subgroup?”, Martin Sulkanen, February 15, 1994, archived May 29, 2023, at https://web.archive.org/

The twenty-five comments preserved on GG are rather mixed. About half of them were immediately in favour, but the concerns of some amateur astronomers are clearly represented, e.g. by two posts by the same female amateur astronomer. One of them maintains:

*"These two posters have summed up my original concerns. As an amateur, I was afraid that, although I *could* read sci.astro.research, everything would be so esoteric that I wouldn't want to. Meanwhile, sci.astro would become nothing but fringe* [...] *the useful middle ground would be lost.* [...] *Replacing "research" with another word – although I can't think of one right now – would help to convey the intent."*[13]

A US professional astrophysicist commented:

*"What could be done would be to moderate sci.astro and create sci.astro.d, the.d standing for "discussion." That would continue to allow the coexistence of amateur and professional discussion but the moderator(s) could filter out the low S/N postings. (I like to think of this idea as a lightly moderated newsgroup, sort of what sci.space has turned into.) And, since the name would be the same, we wouldn't be putting up perceived barriers."*[14]

Another astronomy amateur is in favour:

*"I reckon some moderation would be in order. I subscribed to this group to get useful information, not to wade through reams of* [...] *stuff* [...]. *A lot of people apparently want to learn something from this group, including me, and the only way we can hope to do that is to get more active astronomical/astrophysical professionals to join in and share their thoughts and experiences with us amateurs. Let's do it."*[15]

---

web/20230529164926/https://groups.google.com/g/sci.astro/c/jyxdPTquA-jw/m/5x8x4suBzNEJ, emphasis added.

[13] Ibid. (in the thread), comment dated February 17, 1994.

[14] Ibid. (in the thread), comment dated February 18, 1994.

[15] Ibid. (in the thread), comment dated February 21, 1994.

The first Request for Discussion – the formal step preliminary to vote – came in a long message Sulkanen sent on March 9, 1994.[16] After replicating the February call, he meaningfully added in the end:

> *“Of course, the role of sci.astro is not to provide an exclusive forum for research, however, the present condition of sci.astro greatly inhibits the involvement of the astronomical research community.*
> *Informal discussion about creation of a moderated subgroup of sci.astro has recently appeared in sci.astro and by e-mail.* There was general support for the creation of such a group, provided that it did not adversely affect the interaction between amateur and professional astronomers. *There was particularly strong support for a moderated subgroup voiced by professional and academic astronomers.* […] *The purpose of this newsgroup is the discussion of astronomy & astrophysics research, and the dissemination of information related to astronomy & astrophysics research. Postings appropriate for sci.astro.research would include* […]*: (i) inquiries or discussions about specific current or historical research, or (ii) of research-related topics (observing equipment, computational techniques & software, catalogs, textbooks, journals, references, etc.), (iii) observations of astrophysical phenomena of interest to researchers (novae, supernovae, variable stars, high-energy sources, extragalactic astronomy, planetary astronomy, etc.), (iv) announcement of recent publications submitted to refereed journals or of collections of such publications received as preprints, (v) announcement of*

---

[16] “RFD: sci.astro.research moderated”, Martin Sulkanen. The message was sent to sci.astro (https://web.archive.org/web/20231115170439/https://groups.google.com/g/sci.astro/c/n-wd_QOrP9Q/m/fJZHYvJbqzcJ, archived November 15, 2023), sci.physics, sci.physics.research, sci.space.science and (at least) to two meta-newsgroups: news.groups and news.announce.newgroups. It “may be distributed freely to other relevant newsgroups” (ibid.). Sulkanen writes that the “RFD is being submitted” to the newsgroups listed above “on 28 February 1994”, but a RFD with that date hasn’t been found in any of the newsgroups abovementioned. Instead, on March 25, he informs that “the release of the RFD was not until 10 March 1994” (“RFD: sci.astro.research moderated (repost)”, Martin Sulkanen, archived February 13, 2024, at https://web.archive.org/web/20240213093257/https://groups.google.com/g/sci.astro/c/cwmrlds66Vo/m/noPmlqCF3-wJ).

*future conferences & workshops, proposal or grant announcements of opportunity, and (vi) general scientific news relevant to astronomy & astrophysics."*[17]

As we can see, the range of admitted topics looks rather wide for a research venue, including historical astrophysical research – which one of the commenters didn't appreciate – and topics ii) and iii), which are often popular among amateur astronomers.

The proposed moderator, the same Martin Sulkanen,[18] "will have relatively broad powers", adhering to some notable "basic principles":

*"1. Postings will be judged on their relevancy to scientific research in astronomy and astrophysics.* […] The criterion is *not* the credentials of the author (contributions by amateurs are encouraged), but the relevance of the post to research issues.
*2. Controversial topics and issues in research can be addressed, provided that they are discussed with scientific rigor* […]*; "because I say so" speculations will be redirected to sci.astro.*
*3. "Unverified" astronomical observations will be posted with a disclaimer regarding the reliability of the observation. A verified observation is defined as one that has been checked and is certified for accuracy by the supporting institution* […].
*4. Personal attacks, crossposts irrelevant to astrophysics/astronomy research, commercial advertisements, political discussions, or posts originating from addresses that cannot receive e-mail will be rejected."*[19]

---

[17] "RFD: sci.astro.research moderated", Martin Sulkanen, March 9, 2024, archived November 15, 2023 at https://web.archive.org/web/20231115170439/https://groups.google.com/g/sci.astro/c/n-wd_QOrP9Q/m/fJZHYvJbqzcJ, emphasis added.

[18] Ibid. In fact Sulkanen results to have held this position (e.g.: "sci.astro.research", in "Newsgroups Info Center", archived October 22, 2001, at https://web.archive.org/web/20010224093104/http://www.ibiblio.org/usenet-i/groups-html/sci.astro.research.html).

[19] "RFD: sci.astro.research moderated", Martin Sulkanen, https://web.archive.org/web/20231115170439/https://groups.google.com/g/sci.astro/c/n-wd_QOrP9Q/m/fJZHYvJbqzcJ, emphasis added.

The Call for Vote (CFV) was "planned to be conducted by a Usenet Volunteer Votetaker."[20] In fact there were two calls, posted by Brenda J. Roder (NASA) on 4 and 12 April respectively.[21] The same Brenda Roder announced the results on April 27, 1994.[22] In her words:

> *"Moderated group sci.astro.research passes* […]. *There were 466 YES votes and 15 NO votes, for a total of 481 valid votes. There was 1 abstain and 3 invalid ballots.* […] *There is a five day discussion period* […]. *If no serious allegations of voting irregularities are raised, the moderator of news.announce.newgroups will create the group shortly thereafter."*[23]

It is difficult to say whether about 480 votes represented a wide or limited percentage of the connected astrophysical community, but the situation of the nationalities involved in the voting process is clear enough and shows that a major change had occurred. While net.astro appears to have been entirely US-based, those who voted about creating sci.astro.research were based in different countries. Building on the domain of the email addresses Brenda Roder reports for each of the 481 valid votes, and in addition to a 16.21% of votes whose geographic origin looks uncertain,[24] we find that 18.50% of votes comes from Europe (most of them from Germany and the UK, although Norway, Austria, Holland, Sweden, Switzerland, Finland, Poland and Denmark are represented as well). The US account for a 49.48% (probably underrated, as many of the emails from private companies might be American); other coun-tries account for a further 15.80%.[25] As understandable, the great majority of votes comes from academic and research domains. The very few "no" votes come mainly from the US (73.33%); apparently none from Europe, arguably showing how happy researchers in this area may have been to join international specialistic conversations on astrophysics.

[20] Ibid.

[21] "First call for votes (of 2)" and "Last call for votes (of 2)", Brenda J. Roder, April 4 and April 12 1994 respectively, archived January 21, 2024, at https://web.archive.org/web/20221228111323/https://groups.google.com/g/news.announce.newgroups/c/n-wd_QOrP9Q (in a thread).

[22] Ibid.

[23] Ibid.

[24] As they come from commercial internet providers.

The newsgroup's inaugural message, sent on May 2, 1994, has been preserved ("Welcome to sci.astro.research").[26] Here, Sulkanen reiterates his will to not marginalise amateur astronomers, provided their contributions are sufficiently relevant for research purposes:

> *"Particular attention will be paid by the moderator to include contributions by amateur and professional alike, but requiring that posts be relevant to astronomy/astrophysics research issues* […] *"; "This newsgroup <s.a.r.> is about research activities in astronomy & astrophysics, open not only to the professional researchers,* but to amateurs that are carrying out (or would like to start) research programs *(e.g. variable star observations)* as well."[27]

The founder also clarified that he had envisioned a venue for interactivity:

> *"through Mosaic & the World-Wide Web (WWW) there now exists a great deal of information available for the research astronomer, and the purpose of this newsgroup is not be redundant with those services. Thus,* this newsgroup will probably stress the quasi-interactive aspects of Usenet rather than the archival aspects of the WWW".[28]

---

[25] This presence is largely dominated by Canada and Australia but includes Japan (3 votes) and one vote each from Mexico, Brazil, South Africa, New Zealand and Soviet Union.

[26] "Welcome to sci.astro.research", SAR Moderator (Martin E. Sulkanen), archived January 21, 2024, at https://web.archive.org/web/20221219164151/https://groups.google.com/g/sci.astro.research/c/jtKZVHDWhk8.

[27] Ibid.; emphasis added.

[28] Ibid.; emphasis added.

This seems to resonate with one of William Sebok's aims: the affordance of newsgroups as communication tools, perhaps even more than the potential of the paradigm change, in progressive transition to the 2.0. mode, had been grasped very well.

As Sulkanen had claimed in his call to readers, the creation of sci.astro.research after that of sci.astro parallels what had happened in physics one year before, with the debut of sci.physics.research (February 1993) after sci.physics and with the same motivation: the will to have a moderated group available.[29] Elsewhere, Sulkanen is very clear about this: "sci.physics.research [is] my prototype for s.a.r."[30]

Around 1995, readers of sci.astro.research were estimated at "7300" and messages at 42 per month.[31]

Even after the creation of sci.astro.research, at least one further initiative appeared for sizing down sci.astro further by splitting it into different sub-groups and making some of them moderated (while sci.astro would be renamed to "sci.astro.misc"). The idea was

[29] "Sci.physics.research is a newsgroup intended to facilitate relatively noise-free discussions of issues in and about physics. It grew out of the unmoderated group sci.physics in February 1993 as a response to a perceived signal-to-noise-ratio problem in the unmoderated group, which, it was claimed, had diminished the value of that group to the working physicist [sic] […]" ("Posting to sci.physics.research", n.d., archived May 26, 2000, at https://web.archive.org/web/20000526014307/http://www.math.ucr.edu/home/baez/spr.html).

[30] "Even more about sci.astro.research (longish)", Martin Sulkanen, February 23, 1994, archived November 28, 2023, at https://web.archive.org/web/20231128101657/https://groups.google.com/g/sci.astro/c/sOA9pbUZyNM/m/iUZkqQr6mr8J. In 2001 the then-moderator of sci.astro.research, Martin Hardcastle, would confirm: "sci.astro.research […] should be 'lightly' moderated, *bearing the same relation to sci.astro as sci.physics.research does to sci.physics*." ("Sci.astro.research", September 20, 2001, archived January 12, 2024, at https://web.archive.org/web/20240121165427/https://groups.google.com/g/sci.astro.research/c/5g65B3S6C4Q; emphasis added).

[31] "Sci.astro.research", n.d., in "Newsgroup Info Center", copyrighted 1995 by Kevin Atkinson, archived February 24, 2001, at https://web.archive.org/web/20010224093104/http://www.ibiblio.org/usenet-i/groups-html/sci.astro.research.html.

launched through a "Request for Discussion" message posted on sci.astro on June 21, 1995.[32]

A similar process had occurred in mathematics less than two years before. Even after the creation of sci.math.research, an RFD dated February 15, 1993 stated that "consensus seems to have it [sic] that sci.math should be split" to get a "clear separation of interests between recreational, educational and professional research in mathematics."[33]

Although many commenters agreed about the too high volume of traffic on sci.astro, and the frequent uninteresting posts, only three people were clearly in favour of the proposal.[34]

A modified Request for Discussion was posted on July 20, 1995;[35] the call for vote hasn't been found.

---

[32] "Traffic in sci.astro has reached more than 2000 articles per month (about 80 per day). This is close to the limit where it is recommended to split a group. In addition, there is an increasing number of threads dealing with "non traditional" astronomy, which makes it difficult to follow any kind of discussion. This is so severe that many posts recently complained about the high traffic and noise to signal ratio. One of the main current threads has "cranks" as subject… That aside, although there are a few specialized groups under sci.astro (e.g. fits, hubble, planetarium, research), the subjects addressed in sci.astro are extremely varied (from the Earth to distant galaxies, and from time conventions to extra terrestrial life). Therefore, it appears obvious that sci.astro desperately NEEDS to be split." ("RFD: sci.astro reorganization", Philippe Brieu, June 21, 1995, archived March 22, 2023, at https://web.archive.org/web/20230322111040/https://groups.google.com/g/sci.astro/c/DHOyAPWagg0/m/cl7_Nl4PeMIJ).

[33] "RFD: sci.math.{educational,recreational}", Alex Lopez-Ortiz, February 15, 1993, https://www.usenetarchives.com/view.php?id=news.announce.newgroups&mid=PDFscGM5MUlOTmpqMUByb2Rhbi5VVS5ORVQ%2B, accessed January 10, 2025 (this message hasn't been found on GG).

[34] Around the end of the discussion, a commenter wrote: "The thread on re-organization which ran in news.groups appears to have died. Looks like it's time for the the CFV. Personally, I feel that the RFD didn't even generate enough discussion to warrant a CFV" (June 30, 1995, https://www.usenetarchives.com/view.php?id=sci.astro&mid=PDN0MW5jZiQxNnR, accessed January 10, 2025).

[35] "2nd RFD: sci.astro reorganization", Philippe Brieu, archived January 22, 2024, at https://web.archive.org/web/20240122161944/https://groups.google.com/g/news.announce.newgroups/c/Au_jhdbDUlk/m/ySPqbc765msJ.

## 8. *Elements for a contextual understanding of net.astro, sci.astro and of their specialistic counterparts*

The creation of the two newsgroups examined, in 1983, and their progressive uptake are clearly located in the range of phenomena pertaining the virtualization of communications in research environments.

It may be reckoned that a substantial line of evolution in these experiences revolved around one main issue: the esoteric vs. (to some degree) exoteric nature of these newsgroups – in other words, preferring a resource for professional astrophysicists only vs. opting for one or two venues that be open also to laypeople active on the subject – and the degree of this opening. When sci.astro.research was created in 1994, this contraposition had evolved to that of simply distinguishing high level from low-level interventions, whatever the professional category of the posters was – thus showing that a major shift in perspective had occurred.

Indeed, multiple phenomena were at play in those same years, showing that the traditional boundaries of the academic workforce and – to a much lesser extent – goals and methods were being extended to some degree due to a closer relationship with society. A tentative and non-exaustive list can include:

– a growing involvement of the astrophysics community in social and political issues since the first half of the twentieth century;[1]
– the consideration regained by science amateurs for the making of science since the second half of the same century.[2] In spite of

[1] Jörg Matthias Determann, *Diversity, Equity, and Inclusion in Astronomy. A Modern History* (Springer, 2023), https://doi.org/https://doi.org/10.1007/978-3-031-46113-2. According to Perkmann et al., this aspect of engagement is still underexplored (Markus Perkmann et al., "Academic Engagement: A Review of the Literature 2011-2019," *Research Policy* 50, no. 1 (2021): 104114, https://www.sciencedirect.com/science/article/pii/S004873332030189X).

[2] Arthur J. Meadows, *Communicating research* (Academic Press, 1998), 26-27. Meadows locates a period of decline from the end of the nineteenth century to the first half of the twentieth century, both in astrophysics and within science at large.

a sometimes ill-concealed sense of "obvious" hierarchical subordination of amateur astronomers to professionals,[3] the claim of possible fruitful relationships between proficient amateurs and professional astrophysicists has been found repeatedly also in specific literature around the period examined,[4] thus testifying that the most advanced tier of amateur astronomers could in fact increasingly be considered a source of collaborators for professional astrophysicists.[5] This high opinion of amateurs' potential

---

[3] E.g.: "our own experience of interacting with amateur astronomers has generally been a gratifying and human-enriching one […]. *One must however be prepared to handle an unavoidable fringe of weird and crackpot characters attracted by our science*" (André Heck, "Communicating in Astronomy," in *Organizations and Strategies in Astronomy,* ed. André Heck (Kluwer, 2000), 177, emphasis added). A distinction was made between "an amateur astronomer" and "what I identify […] as a "recreational sky observer" (Thomas R. Williams, "Criteria for Identifying an Astronomer as an Amateur," in *Stargazers. The Contribution of Amateurs to Astronomy, Proceedings of Colloquium 98 of the IAU, June 20-24, 1987*, ed. Storm Dunlop and Michèle Gerbaldi (Springer, 1988), 24).

[4] Robert A. Stebbins, "Looking Downwards – Sociological Images of the Vocation and Avocation of Astronomy," *Journal of the Royal Astronomical Society of Canada* 75 (1981): 2-14; Robert A. Stebbins, "Amateur and Professional Astronomers: A Study of Their Interrelationships," *Urban Life* 10, no. 4 (1982): 433-54; Robert A. Stebbins, "Amateurs and Their Place in Professional Science," in *New Generation Small Telescopes*, ed. D.S. Hayes, D.R. Genet, and R.M. Genet (Fairborn Press, 1987), 217-25. From an amateur's perspective: Paul Boltwood, "An Amateur Astronomer's Experiences with Amateur-Professional Relations," in *Amateur-Professional Partnerships in Astronomy (ASP Conference Series, 220)*, ed. J.R. Percy and B. Wilson (Astronomical Society of the Pacific, 2000), 188-95, https://adsabs.harvard.edu/pdf/2000ASPC..220..188B.

[5] This reminds us of the previously cited statement by Bill Sebok: "Amateur astronomers and professional astronomers are mostly (*but not always*) different people" ("Net.astro", November 14, 1983, in a thread, emphasis added; archived January 13, 2024, at https://web.archive.org/web/20240131142016/https://groups.google.com/g/net.news.group/c/76iybyGt5uQ). Also, as US astrophysicist Michael Turner put it on net.news.group in the same days, "to make a distinction between a net.astro and a net.astro.amateur is not a good one, considering the very significant contributions of the amateurs in the field. The popular view of astronomy is considerably out of proportion to its funding as a science, and to its professional population--about 2000 researchers in the whole world, I think,

contribution is at the heart of the rationale for sci.astro.research, as we have seen;

– the growing role taken approximately since the mid-1990s by the outreach activities, enacted both by research institutions and by single researchers or small groups.[6] A prospective development would be the citizen-science projects successfully launched since the second half of the same decade;[7]
– the ever-increasing role played by technology in contemporary astrophysics.[8] This process, one of broad scope in science,[9] seems to be remodelling a significant part of the discipline into a mul-many of whom are not paid full-time. A great deal of valuable research is carried out by astronomers who are not paid at all." (November 18, 1983, in a thread; archived December 5, 2023, at https://web.archive.org/web/20231205101624/https://groups.google.com/g/net.news.group/c/coqAzLPeKAU). In the same vein, another newsgroup post sent by US astrophysicist Robert W. Spiker on March 22, 1994 claimed that "much data on novae and variable stars are taken by what are normally termed amateur astronomers" (in a thread; archived January 15, 2024, at https://web.archive.org/web/20240115145345/https://groups.google.com/g/news.groups/c/Ii9SXTfiDZs/m/lAciRUrnXAgJ).

[6] For the US, e.g.: Kenneth Scott Edgett et al., "K-12 and Public Outreach for NASA Flight Projects: Five Years (1992-1997) of the Arizona Mars K-12 Education Program," in *28th Annual Lunar and Planetary Science Conference* (Houston, 1997), 323-24, https://www.lpi.usra.edu/meetings/lpsc97/pdf/1124.PDF; Jeffrey D. Rosendhal et al., "The NASA Office of Space Science Education and Public Outreach Program," *Advances in Space Research* 34, no. 10 (2004): 2127-35; Lynn R. Cominsky, "Education and Public Outreach in Astronomy and Beyond," *Nature Astronomy* 2 (2018): 14-15, https://doi.org/10.1038/s41550-017-0359-y.

[7] About citizien science at large: Bruno J. Strasser et al., "'Citizen Science'? Rethinking Science and Public Participation," *Science & Technology Studies* 32, no.2 (2019): 52-76, https://doi.org/10.23987/sts.60425; in astrophysics: Carol Christian et al., 2012. "Citizen Science: Contributions to Astronomy Research," https://arxiv.org/abs/1202.2577.

[8] E.g.: Karin Knorr-Cetina, *Epistemic Cultures*, 27; Rocky Kolb, "A Thousand Invisible Cords Binding Astronomy and High-Energy Physics," *Reports on Progress in Physics* 70, no. 10 (2007): 1583-95, https://doi.org/10.1088/0034-4885/70/10/R01.

[9] Alfred Nordmann, Hans Radder, and Gregor Schiemann, eds., *Science Transformed? Debating Claims of an Epochal Break* (University of Pittsburgh Press, 2011).

tidisciplinary and large-team effort, blurring the borders between scientific and technological or technical activity and implying that interprofessional and/or interdisciplinary communication is more and more usual in the everyday practice of astrophysics.

Under this scenario, scholarly communication on a less restricted professional basis might be seen as a necessary evolution of a previously elitist communication scheme, and an increasing amount of diversity[10] among the actors of communication may be considered as naturally related.

Connectedly, the internet has been regarded as a specially suited setting for the encounter of diverse people, as it "allows relationships to develop on the basis of shared interests rather than be stunted at the onset by differences in social status".[11] This seems to have been particularly true at the beginning of the 1990s, after the much larger diffusion of the internet and the related, worldwide and generally optimistic view of its "democratizing" role.[12]

[10] About the role of diversity in science teams see Hall, Kara L. et al., "The Science of Team Science: A Review of the Empirical Evidence and Research Gaps on Collaboration in Science," *American Psychologist* 73, no. 4 (2018): 536. Also: Richard Freeman and Wei Huang, "Collaboration: Strength in diversity", *Nature* 513, no. 305 (2014), https://doi.org/10.1038/513305a; More generally: Andreas Hundschell et al., "The Effects of Diversity on Creativity: A Literature Review and Synthesis", *Applied Psychology* 71 (2022): 1598-634.

[11] Barry Wellman and Milena Gulia, "Virtual Communities as Communities: Net Surfers Don't Ride Alone," in *Communities in Cyberspace*, ed. Peter Kollock and Marc Smith (Routledge, 1999), 186. As it has been noted, "where the distinction between professionals and amateurs was once clear, the adoption of online services by each puts these distinctions into question" (Liz Dowthwaite and James Sprinks, "Citizen science and the professional-amateur divide: lessons from differing online practices", *Journal of Science Communication* 18, no.1 (2019), A06, https://doi.org/10.22323/2.18010206. As for an increased involvement of students and "non-elites scientists" in the context of a still elite-dominated scientific environment, see Finholt, Thomas A. and Gary M. Olson, "From Laboratories to Collaboratories: A New Organizational Form for Scientific Collaboration," *Psychological Science* 8, no. 1 (1997): 34-35, https://www.jstor.org/stable/40062842?read-now=1&seq=7#page_scan_tab_contents.

Thus, a historical framing of our experiences based on the coupling of the two major phenomena mentioned above – wider connection between science and society and diffusion of the internet in social groups connected to science – may perhaps fit our experiences adequately. Nevertheless, this operation would require a comprehensive sociological or philosophical treatment that is outside the purpose of the present research. We won't follow this ambitious path. As Peters insighfully put it, "while philosophy, sociology, and history continue to grapple with metaphysically large framing questions, perhaps the framing questions of communication research that takes its history seriously carry with them a certain strength of its own modesty […] In other words, […] *the* most basic question remains the most vital: how, exactly do people use machines to use people? To repeat Norbert Wiener's take on postwar communication research, it is not just about the machines: communication is about the human use of human beings".[13] Following this approach, some elements will be noted to improve our understanding of what a satisfactory newsgroup experience was expected to be and was perceived to be by the astrophysicists concerned. This understanding will thus be guided from the feedback of the community, in a grassroots and contextual rather than conceptual perspective. It is our belief that this approach may be fruitful also from a socio-scientific point of view.

[12] Michael Hauben, "The Computer as a Democratizer", in Michael Hauben and Ronda Hauben, *Netizens: On the History and Impact of Usenet and the Internet* (IEEE Computer Society Press, 1997), chap. 18, http://www.columbia.edu/~rh120/ch106.x18 (accessed 26 Nov 2024); Emily B. Laidlaw, "The Internet as a Democratising Force," in *Regulating Speech in Cyberspace: Gatekeepers, Human Rights and Corporate Responsibility* (Cambridge University Press, 2015), 3; James Curran, "Rethinking Internet History", 35.

[13] Benjamin Peters, "On Digital Media History," *Journal of Communication* 74, no. 6 (2024): 510, https://doi.org/https://doi.org/10.1093/joc/jqae028.

As a first question, we can ask ourselves if the mixed professional status of the participants to our newsgroups – especially after the beginning of the nineties, where participating to the online communication arena was much more common in the Western world – made our scholarly newsgroups a success.[14] Could professionally "mixed" astrophysical newsgroups take advantage of the stimuli coming from an extended "communication audience" and really become the venue(s) for the lively scholarly debates which William Sebok and Martin Sulkanen had hoped for in 1983 and 1994? Did these newsgroups happen to be the cradles of new ideas?

Based on our data, it must preliminarly be acknowledged that the astrophysicists who joined Usenet newsgroups specific for the discipline (whether moderated or unmoderated; whether general or specialist in scope) have been a minority. One of the individually interviewed astrophysicists, asked whether it was common to use astrophysical Usenet newsgroups in his working environment in the USA, answered: "Not really. Even in the time of sci.astro and sci.astro.research, professional astronomers already had ways of exchanging preprints of early research and meeting at conferences. Thus, while some professional astronomers did participate in sci.astro and sci.astro.research, but it was not many."[15]

The researchers who answered the questionnaire administered within the Italian National Institute of Astrophysics revealed that those who used Usenet newsgroups specific to the discipline were 16.52%.

Why had these researchers decided to join astrophysics newgroups?

According to the results of the questionnaire, the main reasons were two, almost equally represented: on the one side, "faster sharing of information" (43.10%), and on the other side the process which leads to new ideas: "possibility to get to new ideas" (24.13%) and "possibility to get involved in interesting debates" (18.96%), which may be merged on a subject basis, so that, altogether, they account for a 43.09%. These motivations are significantly aligned with those found by Berge and Collins for scholarly communication at large,[16] and, especially, correspond to the motivations expressed by the founders for creating (or modifying) our two newsgroups.

[14] The issue of the success of internet newsgroups at large has been interestingly debated by Marc A. Smith, "Invisible Crowds in Cyberspace". This paper, though, analyses different features than those considered in the present paragraph.

[15] C1.

According to our data, at least the first goal – information sharing – seems to have been achieved in an overall satisfactory manner. About sci.astro, unmoderated, one of our informants recalls:

> *"I think that there were three main uses. First, there were people who wanted to learn more about astronomy. Typically, these were fairly basic questions or misconceptions, such as the Earth being closer to the Sun in December rather than July. The second was sharing of recent results.* […] *The final set of people were those who had their favorite model of astronomy or of the Universe to advertise* […]*."*[17]

According to another individual interviewee, who reported his experience in an academic institution in Germany,

> *"by 1992 both usenet and email played a role in professional communication in astronomy, but were still dominated by telephone, fax, and conventional letters.* […] *At that time, I would say that email and Usenet were about equally relevant, but used for different things.* […] *FORMAL communication was still by paper."*[18]

---

[16] "Mature scholarship typically includes the generation and analysis of ideas, research and theory building, and the evaluation and dissemination of the results of these activities. In general, persons engaged in discussing scholarly activity are motivated by a mutual interest. *Their goals are to inform themselves or others and to help problem-solve through scholarly communication.*" (Zane L. Berge and Mauri Collins, "Computer-mediated scholarly discussion groups", *Computers & Education* 24, no. 3 (1995): 184; emphasis added).

[17] C1.

[18] C2.

In the questionnaire created for the present research, when asked if newsgroups were a good place for the exchange of information, as many as 67.57% of the respondents said they "agree" or "very much agree" (51.35% and 16.22% respectively).

The problem with astrophysics newsgroups, then, seems to have to do (or, better, have had, as 89.19% of the respondents to the survey report they have now stopped using them) with the second goal, the "possibility of getting to new ideas" and "to get involved in interesting debates." This is confirmed by the largely prevailing answer to the question about why they had stopped using astrophysics newsgroups: 68.42% thought they were "not interesting enough."[19]

About sci.astro.research, which is moderated, an individual informant reveals that his personal experience was different than expected:

*"I participated in sci.astro.research only as a passive user; in fact I thought it would be about talking about new projects, but mostly it was dissemination of news* […] *the communication taking place there was mainly informative."*[20]

At this point, we would like to introduce the take by Bruce Lewenstein, who examined the newsgroup sci.physics.fusion – about nuclear physics – and the behaviours observed there during the "cold fusion saga" (1989-1992). Far from simply reporting what happened on the newsgroup, Lewenstein cast light on the relationship between information availability and the social making of science on a CMC tool. As he put it, "knowledge […] would develop as people came to consensus about the big ideas, as they agreed which big ideas were no longer isolated pieces of information but […] more fully developed ideas to which others could subscribe as established knowledge."[21] In his view, "the impact of these new technologies", including newsgroups, "was largely confined […] to issues of awareness and information gathering".[22] It is true that "they accelerated the speed at which individuals had to respond to new ideas and new information, and they affected who had access to what information at what time"[23]; also, notably, they "*entered into the process by which researchers made judgments about the new, fast-moving* […] *area* […] *and thus were part of the process by which social consensus – knowledge – was produced.*"[24] Nevertheless, he stresses that the process of transitioning from information availability to the making of new ideas can be impaired by a too high level of "noise,"[25] which he estimates at 30% in the newsgroup he had analyzed.[26]

[19] A further 26.32% reported they had "not enough time"; for 5.26% there was "too much spam".

[20] C3 (English translation by the author).

[21] Lewenstein, "Do Public Electronic Bulletin Boards", 132.

Lewenstein puts the high level of "noise" encountered in the scientific newsgroup examined in close relation to the composite nature of its public. The latter, he claims, included both "participants" (i.e., professional scientists) and "the nonprofessional observers," the former being the minority.[27] Allegedly, as a consequence, "because so many active users of the newsgroup were not themselves experts in the fields under discussion, misinformation was common."[28]

Hence his conclusion:

[22] Lewenstein, "Do Public Electronic Bulletin Boards", 142.

[23] Lewenstein, "Do Public Electronic Bulletin Boards", 124.

[24] Lewenstein, "Do Public Electronic Bulletin Boards", 142; emphasis added.

[25] "[Most] active researchers followed the net only rarely (perhaps a few times a year), if at all. *Most likely, they found the "signal-to-noise ratio" of the public newsgroups too poor for the nets to be useful"*. (Lewenstein, "Do Public Electronic Bulletin Boards", 143; emphasis added).

[26] Lewenstein, "Do Public Electronic Bulletin Boards", 138.

[27] "The vast majority of the contributors were certainly *not* professionals in the field. […] A […] survey in 1990 found […] that nearly 20 percent of the people who responded (9 out of 49) described themselves as cold fusion experimenters." (Lewenstein, "Do Public Electronic Bulletin Boards", 135-36).

[28] Lewenstein, "Do Public Electronic Bulletin Boards", 140.

*"The role of the electronic bulletin boards*[29] *suggests that they, as completely public forums, did not serve the needs of the active research community.* […] *The evidence suggests that [most] active researchers followed the net only rarely (perhaps a few times a year), if at all. Most likely, they found the "signal-to-noise ratio" of the public newsgroups too poor for the nets to be useful. Thus the nets were most useful for the nonprofessional observers, the people who wanted to know what was going on in cold fusion without themselves taking an active part in the day-to-day work* […]*,"*[30] and, perhaps somehow abruptly,

*"these people [the "nonprofessional observers"] were not the relevant community for making the judgments about which information was most useful for creating stable knowledge."*[31]

A part of the astrophysical community which has been interviewed for the present research seems to be aligned with this appraisal, if we rely on some optional comments. Very clearly: "Usenet newsgroups [are] not necessary. Fast circulations does not necessarily mean good ideas […]"; newsgroups are "more suited to chat-like or quick request of information than to science support"; and, importantly, they are "good for information spreading, but [there are] too many heads for idea convergence."

Other critical comments labelled astrophysical Usenet newsgroups as "often not focused and used more 'just to show to be there'", or, even worse, "useless". An apparently traditionally-oriented researcher noted that "there was no value in debating on usenet. The real debate is done with scientific papers."

About the unmoderated sci.astro, one of the individually interviewed astrophysicists is particularly clear, referring to the news-group as available around the beginning of the nineties: "I am not sure that sci.astro was ever that helpful to the broader astrophysical community. There were certainly some professional astronomers who participated, but the group was open to all and there were certainly a number of quite basic questions asked or topics that seemed to appear on a regular basis."[32]

[29] Lewenstein uses the expressions "electronic bullettin boards" and "Usenet newsgroups" as synonims.

[30] Lewenstein, "Do Public Electronic Bulletin Boards", 143.

[31] Ibid. (emphasis added).

These comments seem to resonate with the opinion expressed by physicist Paul Ginsparg about his own discipline in 1994: "Usenet newsgroups, for reasons such as their lack of indexing and archiving, and excessively open nature, are unlikely to prove adequate for serious purposes," at least as intended as "scholarly research communication."[33]

In fact, though, Lewenstein acknowledges that things may be different in other scientific domains, and even in some sub-domains of physics itself: "*other bulletin boards systems (including ones for* […] *high-energy physicists* […]*) may be more useful to researchers.*"[34]

As for astrophysics, one of the individually interviewed researchers, asked if he thought sci.astro.research had been closer to the needs of the astrophysical community than sci.astro, answered it might have been.[35]

A positive appraisal has been found in the literature, in the same 1994, although it's referred to newsgroups' "static" working mode.[36] Also on the positive side, an optional comment from one of the respondents to the questionnaire stated that Usenet newsgroups "worked fine. Simply, we have now more possibilities."

[32] C1.

[33] Ginsparg, "First steps," 395.

[34] Lewenstein, "Do Public Electronic Bulletin Boards", 143; emphasis added. As noted above, this author uses the expression "bullettin board system" and the noun "newsgroup" as synonyms.

[35] (Q.) "Do you think sci.astro.research was closer to the needs of the astrophysical community [than sci.astro]?" (A.) "I think so, but, by the time that it was created, I already may have been getting busy enough with the late stages of my graduate school and into my postdoctoral position that I was participating less in it." (C1).

[36] "A nice spinoff of Usenet newsgroups is that many individuals have taken it upon themselves to maintain FAQ (Frequently Asked Questions) files. […] As a condensation of the collective wisdom of the newsgroup (and admittedly signal-

The same questionnaire reveals that newsgroups are considered to be a good place for communicating on research subjects (59.46% of answers: 43.24% "agree", plus 16.22% "agree very much").[37] As important as it seems, this result should probably be taken cautiously, as we don't know to what extent these answers referred to unmoderated or to moderated newsgroups – or to very specialistic sub-newsgroups (possibly, external to the ones under examination), which cut down non-specialists somehow automatically.[38] This interpretation may be confirmed by the very high percentage of respondents who thought the newsgroups they had experienced included few spammers and few people who breached the netiquette.[39] Secondly, some respondents might have misinterpreted the action of "communicating" as the simpler task of disseminating information – a difference which Martin Sulkanen had marked clearly.

Anyway, the scholars surveyed have largely acknowledged the feature of diversity in their experience of astrophysics newsgroups: actually, the latter "allowed the presence of multiple points of view" according to as many as 80%.[40]

---

to-noise often leaves much to be desired), FAQs provide excellent all-around introductions to a topic, and comprehensive background information." (Andernach, Hanisch, and Murtagh, "Network resources for astronomers," 1191).

[37] 27.03% "don't agree very much"; only 5.41% "disagree."

[38] One of these sub-newsgroups was sci.astro.fits, which some respondents explicity mentioned in their optional comments.

[39] Q.: "How much do you agree with the following: 'Too many people breached the netiquette or were spammers?'. "I don't agree very much/ I disagree / I very much disagree" altogether account for 70.59% of the respondents (in detail: "I disagree", 23.53%; "I don't agree vey much", 44.12%; "I very much disagree", 2.94%).

[40] Sum of the 74.29% of respondents who chose "I agree" plus the 5.71% who "very much agree". 17.14 % "don't agree very much"; 0% "disagree"; 2.86% "very much disagree".

Diversity has also been reported by one of the individually interviewed astrophysicists, with specific reference to sci.astro:

> *"the audience of sci.astro was diverse. There were some professional astronomers and astronomy graduate students, there were amateur astronomers, people interested in astronomy, and some people whose participation was not always helpful."*[41]

Now, diversity has been acknowledged as a prominent feature of CMC settings, one that enhances online communication in specialistic environments and, thus, a possible harbinger of their success.[42] It is our belief that reporting diversity in these CMC tools may be interpreted as an implicit mark of appreciation, although generalizations would be erroneous.

Some further optional comments pointed out that the reasons for stopping using Usenet newsgroups were contingent.[43]

On the whole, we substantially speculate that this community's take about disciplinary Usenet newsgroups is generally more nuanced than it has been reported to be for the users of sci.physics.fusion. On the basis of the present analysis, it seems safe to assume that a satisfactory scholarly experience with Usenet newsgroup in the domain of astrophysics is deemed to have occurred, thanks to a balance of factors including number and type of participants, mode of their participation[44] and scope of the newsgroup. As it has very well been synthesized in general terms, "many discussion forums, be they mailing lists, web-based discussions, or IRC channels, pro-duce high-quality discussions among a few participants. The challenge is to scale this "few-to-few" communication all the way to "many to many". Adding a moderator helps to a point. […] Rarely can a discussion forum have enough commonallity in interest to draw a critical mass of audience, have enough variety in viewpoints to keep the discussion interesting, and scale to ever-larger audience sizes without lossing [sic] the signal of the discussion in the noise of the chatter."[45]

---

[41] C1.

[42] Paloque-Bergès, *Qu'est-ce qu'un forum internet?*, 73.

[43] "They were temporary newsgroups"; they "were discontinued in favour of mailing list"; "most of them disappeared." Also, "social media groups are more effective."

[44] About the number of participants in relation to the success of newsgroups, Smith maintains that "newsgroups [which] attract between fifty and fifty-hundred people […] may be the most productive and stable of all newsgroups." (Smith, "Invisible Crowds in Cyberspace," 12).

A final element, which we consider to be relevant for our users' positive appraisal of newsgroups, be it entirely aware or not, is the specific fruitfulness of this experience for the generation of new ideas.

Our questionnaire included the question: "in your experience, have new ideas ever been stimulated by interactions on astrophysics newsgroups?" Interestingly, the majority of the answers was "yes" (47.05%), and even more interestingly 29.41% specified "yes, but then the ideas were discussed in other ways (in person, by email…)."[46]

The practice of using different communication tools and/or settings at the same time, and cross-posting parts of the discussion(s) from one to another CMC tool or ecosystem looks frequent: it is attested, for example, in the newsgroup post Bill Sebok sent about the email comments he had received about the creation of net.astro,[47] or by Martin Sulkanen on sci.astro, in the thread about the launch of sci.astro.research.[48] After one of her posts on sci.astro, amateur astronomer Brenda Kalt reports that "several people have e-mailed me, and I thank them […]".[49]

---

[45] Jack Bates, and Mark Stone, "Communicating Many to Many," in *Open Sources 2.0: The Continuing Evolution*, ed. Chris DiBona, Danese Cooper, and Mark Stone (O'Reilly, 2005), 379-80, https://www.oreilly.com/library/view/open-sources-20/0596008023/.

[46] In detail: "yes, with output on the newsgroup": 17.64 %; "yes, but then the ideas were discussed in other ways (in person, by email…)": 29.41%; "I don't know": 23.52%; "I don't remember": 3.92%; "no": 25.49%.

[47] "For those of you who have attempted to reply to me on this topic to xanth.msfc.nasa.gov, my apologies for bouncing mail […] I'm now taking mail on the subject […]. I've received some very thoughtful constructive comments, and I'll be posting an abridged version of them pretty soon." ("Summary of net.astro responses", Bill Sebok, November 21, 1983, archived April 4, 2024, at https://

This is in line with Lewenstein's remark: "*It is clear* […] *that many people read the net and contributed ideas without directly posting messages to the net,"*[50] or also, many people read CMC tools and contributed ideas to the net *without necessarily using the same CMC tool*. This phenomenon may have to do with that of lurking,[51] but may also be convincingly framed by the take by Wellman and Gulia for which "newsgroups and discussion lists provide permeable, shifting sets of participants, with more intense relationships continued by private email."[52]

Substantially, from the point of view of newsgroup research in scholarly environments, it may be misleading to decide about the usefulness of these CMC tools by judging *exclusively* on the new ideas expressed and developed on them. Cleverly, Lewenstein was aware of this problem: "to answer this question fully requires a

---

web.archive.org/web/20230404121504/https://utzoo.superglobalmegacorp.com/usenet/news007f3/b19/net/news/group/866.txt).

[48] "Is it time for a moderated research subgroup?", Martin Sulkanen, February 15, 1994, archived November 15, 2024, at https://web.archive.org/web/20240115161129/https://groups.google.com/g/sci.astro/c/jyxdPTquAjw/m/5x8x4suBzNEJ (with threaded comments).

[49] Ibid., in the thread.

[50] Lewenstein, "Do Public Electronic Bulletin Boards", 135.

[51] "It seems safe to assume that more people read the Usenet than actively participate in it. […] Newsgroups have many silent observers for every active poster" (Smith, "Invisible Crowds in Cyberspace", 4).

[52] Wellman and Gulia, "Virtual Communities," 181. At a later stage, after the social media were born, Rowlands et al. have maintained that researchers who are "active social media users […] disseminat*e* their findings through email lists and Web groups, personal web-pages, wikis, blogs, social networks, and Twitter." (Rowlands, Ian et al., "Social Media Use in the Research Workflow," *Learned Publishing* 24, no. 3 (2011): 192, https://doi.org/10.1087/20110306).

detailed analysis of all the forms of communication used in a particular context."[53]

The astrophysicists interviewed through the survey seem to have implicitly suggested that an excessive focus on specific CMC tools for localizing the making of new ideas may result to be artificial, and certainly answered that new ideas in fact happened to be stimulated by posts on newsgroups, and then discussed elsewhere. For as much as the present research is concerned, we believe that this information can be considered as a clue for acknowledging Usenet newsgroups a role as possible transmission gears for the building of new ideas, rather than seeing them as a "walled" venue of discussion whose relevance may be weighted exactly. Also, these users appear to have been pragmatic enough to withdraw from this CMC tool later on, when its role had declined ("not interesting enough," i.e. having too little to leverage on for acting as stimuli).

The uptake of specialized newsgroups by a part of the scholarly astrophysical community starting from the 1980s witnesses that a need was in the air and that experimenting with this new online communication tool was deemed to be potentially useful. Also, the public conceived at least for sci.astro.research by its creator in 1994 testifies that the idea of a "common room" for professionals and for some proficient non-professionals in the name of astrophysics was desirable and seemed to be feasible, although, as it has been noted, "addressing laypeople and its implications in the [digital] scholarly spaces have always been complicated."[54]

Finally, the astrophysics community has shown

---

[53] Lewenstein, "Do Public Electronic Bulletin Boards," 124. The question is "what effects those new patterns had on the process by which isolated bits of information […] became robust contributions to 'scientific knowledge', supported by the social consensus that produces stable knowledge."

[54] ("[…] l'adresse au profane et son implication dans les espaces savants ont toujours été compliquées." (Paloque-Bergès, *Qu'est-ce qu'un forum internet?*, 96; English translation by the author).

– the capacity to build a new CMC tool for professional communication on its own, which will be replicated little time later for ArXiv-based commenting resources;[55]
– a constant effort to self-regulate its major Usenet newsgroups by resizing broadness of scope and diversity of admitted participants, in order to optimize the appeal and fruitfulness of these tools for the community itself.

As it has been noted, only an analysis of the entire disciplinary communication ecosystem, as complex as it would be, might cast substantial light on the quality and dynamics of a multi-layered, more and more digital scholarly communication ecosystem after the second third of the 20th century. Nevertheless, it seems to be a fact that the astrophysics community built new gateways for the (conditioned) admittance of different types of socio-professional contributions, influences, or "echoes".

[55] Monica Marra, "Arxiv-based commenting resources by and for astrophysicists and physicists: an initial survey", in *Expanding Perspectives on Open Science: Communities, Cultures and Diversity in Concepts and Practices*, ed. Leslie Chan and Fernando Loizides (IoS Press, 2017), 100-117, https://doi.org/10.3233/978-1-61499-769-6-100.

# *Bibliography*

Abbate, Janet. *Inventing the Internet.* MIT Press, 1999.

Andernach, Heinz, Robert J. Hanisch, and Fionn Murtagh. "Network Resources for Astronomers." *Publications of the Astronomical Society of the Pacific (PASP)* 106, (1994): 1190-216. https://iopscience.iop.org/article/10.1086/133497/pdf.

Andernach, Heinz. "On–line Data and Information in Astronomy or Where to Find Astronomical Information without Scanning the Bookshelves of the Library." *IAC Technical Note*, no. 1 (1993): 1-18. https://articles.adsabs.harvard.edu/pdf/1993IACTN........1A.

Barjak, Franz. "The Role of the Internet in Informal Scholarly Communication." *Journal of the American Society for Information Science and Technology* 57, no. 10 (2006): 1350-67. https://ils.unc.edu/courses/2014_fall/inls690_109/Readings/Barjak2006-RoleOfInternet.pdf.

Bates, Jack, and Mark Stone. "Communicating Many to Many." In *Open Sources 2.0: The Continuing Evolution*, edited by Chris DiBona, Danese Cooper, and Mark Stone. O'Reilly, 2005.

Baumann, Ryan. "Early Usenet History and Archiving." */etc* (blogpost). Originally published February 23, 2015. Archived January 29, 2024, at https://web.archive.org/web/20240129120207/https://ryanfb.xyz/etc/2015/02/23/early_usenet_history_and_archiving.html.

Baym, Nancy K. "From Practice to Culture on Usenet." *The Sociological Review* 42, no. 1 (1994): 29-52.

Becher, Tony, and Paul Trowler. *Academic Tribes And Territories: Intellectual Enquiry and the Culture of Disciplines*. 2.ed. The Society for Research into Higher Education and Open University Press, 2001.

Bellovin, Steve. "The Early History of Usenet. Part VI: The Public Announcement." *CircleID*, November 27, 2019. Archived January 23, 2024, at https://web.archive.org/web/20240123130936/https://circleid.com/posts/20191127_the_early_history_of_usenet_part_vi_the_public_announcement.

Berge, Zane L., and Mauri Collins. "Computer-mediated scholarly discussion groups." *Computers & Education* 24, no. 3 (1995): 183-89.

Boltwood, Paul. "An Amateur Astronomer's Experiences with Amateur-Professional Relations." In *Amateur-Professional Partnerships in Astronomy (ASP Conference Series, 220)*, edited by J.R. Percy and B.

Wilson. Astronomical Society of the Pacific, 2000. https://adsabs.harvard.edu/pdf/2000ASPC..220..188B.

Bumgarner, Lee S. "The Great Renaming FAQ. Part. 1" N.d. (between 1994 and 2006). Archived July 9, 2008, at https://web.archive.org/web/20080709061430/http://danflood.com/cs-content/cshistory/csh_greatrename1.html.

Christian, Carol, Chris Lintott, Arfon Smith, Lucy Fortson, and Steven Bamford. "Citizen Science: Contributions to Astronomy Research." Preprint, ArXiv, February 12, 2012. https://arxiv.org/abs/1202.2577.

Cole, Samantha. "2.1 Million of the Oldest Internet Posts Are Now Online for Anyone to Read." *Vice*, October 13, 2020, archived August 10, 2022, at https://web.archive.org/web/20220810144946/https://www.vice.com/en/article/pky7km/usenet-archive-utzoo-online.

Cominsky, Lynn R. "Education and Public Outreach in Astronomy and Beyond." *Nature Astronomy* 2 (2018): 14-15. https://www.nature.com/articles/s41550-017-0359-y.

Curran, James. "Rethinking Internet History." In *Misunderstanding the Internet*, by James Curran, Natalie Fenton, and Des Freedman. Routledge, 2012.

Determann, Jörg Matthias. *Diversity, Equity, and Inclusion in Astronomy. A Modern History.* Springer, 2023. https://doi.org/https://doi.org/10.1007/978-3-031-46113-2.

Dowthwaite, Liz, and James Sprinks. "Citizen Science and the Professional-amateur Divide: Lessons from Differing Online Practices." *Journal of Science Communication* 18, no.1 (2019), A06. https://doi.org/10.22323/2.18010206.

Driscoll, Kevin. *The Modem World. A Prehistory of Social Media*. Yale University Press, 2022.

Driscoll, Kevin, and Camille Paloque-Bergès. "Searching for Missing 'Net Histories'."*Internet Histories* 1, no. 1-2 (2017): 47-59. https://doi.org/10.1080/24701475.2017.1307541.

Edgett, Kenneth Scott, et al. "K-12 and Public Outreach for NASA Flight Projects: Five Years (1992-1997) of the Arizona Mars K-12 Education Program." In *28th Annual Lunar and Planetary Science Conference*. Houston, 1997. https://www.lpi.usra.edu/meetings/lpsc97/pdf/1124.PDF

Egret, Daniel, Robert J. Hanisch, and Fionn Murtag. "Search and Discovery Tools for Astronomical On-Line Resources and Services." *Astron-*

*omy and Astrophysics Supplement Series* 143, no. 1 (2000): 137-43. https://aas.aanda.org/articles/aas/pdf/2000/07/ds1834.pdf.

Finholt, Thomas A., and Gary M. Olson. "From Laboratories to Collaboratories: A New Organizational Form for Scientific Collaboration." *Psychological Science* 8, no. 1 (1997): 28-36.

Freeman, Richard, and Wei Huang. "Collaboration: Strength in Diversity." *Nature* 513, no. 305 (2014). https://doi.org/10.1038/513305a.

Ginsparg, Paul. "First Steps Towards Electronic Research Communication." *Computers in Physics* 8, no. 4 (1994): 390-96.

Hall, Kara L., et al.,"The Science of Team Science: A Review of the Empirical Evidence and Research Gaps on Collaboration in Science." *American Psychologist* 73, no. 4 (2018): 532-48.

Hanisch, Robert J. "Services Available on the Network." *AAS Newsletter*, no. 62 (1992). https://aas.org/publications/aas-newsletter/nl62/epubsup.

Hanisch, Robert J. "Douglas Tody (1952-2022)." *Bulletin of the AAS* 54, no. 1 (2022): 1-5. https://web.archive.org/web/20240131151737/https://baas.aas.org/pub/2022i034/release/1.

Hardy, Henry Edward. "The History of the Net." Master's Thesis, Grand Valley State University, MI, USA, 1993. Archived November 26, 2024, at https://web.archive.org/web/20241126145904/http://www.devin.com/cruft/hardy.html.

Harley, Diane, Sophia Krzys Acord, Sarah Earl-Novell, Shannon Lawrence, and C. Judson King. "Astrophysics Case Study." In *Assessing the Future Landscape of Scholarly Communication: An Exploration of Faculty Values and Needs in Seven Disciplines*, by Diane Harley, Sophia Krzys Acord, Sarah Earl-Novell, Shannon Lawrence, and C. Judson King. UC Berkeley, Center for Studies in Higher Education, 2010. https://escholarship.org/uc/item/15x7385g#page=145.

Hauben, Michael. "The Computer as a Democratizer." (Draft chapter). In Michael Hauben and Ronda Hauben. *Netizens: On the History and Impact of Usenet and the Internet*. IEEE Computer Society Press, 1997. Chapter 18, http://www.columbia.edu/~rh120/ch106.x18.

Hauben, Michael. "The Social Forces Behind the Development of Usenet." (Draft chapter). In Michael Hauben and Ronda Hauben. *Netizens: On the History and Impact of Usenet and the Internet*. IEEE Computer Society Press, 1997. Chapter 3, http://www.columbia.edu/~hauben/book/ch106.x03.

Hauben, Ronda. "On the Early Days of Usenet: The Roots of the Cooperative Online Culture." In Michael Hauben and Ronda Hauben. *Netizens: On the History and Impact of Usenet and the Internet*. IEEE Computer Society Press, 1997. Chapter 10, https://www.columbia.edu/~rh120/ch106.x10.

Hauben, Ronda. "Commodifying Usenet and the Usenet Archive or Continuing the Online Cooperative Usenet Culture?" *Science Studies* 15, no. 1 (2002): 61-68, https://citeseerx.ist.psu.edu/viewdoc/download;jsessionid=B2D0D90EFE821D600B4304A3C51C9157?doi=10.1.1.468.8369&rep=rep1&type=pdf.

Heck, André. "Communicating in Astronomy." In *Organizations and Strategies in Astronomy,* edited by André Heck. Kluwer, 2000.

Heck, André and Fionn Murtagh, eds., *Intelligent Information Retrieval: The Case of Astronomy and Related Space Sciences*, Springer, 1993.

Hocquet, Alexandre, and Frédéric Wieber. "Mailing List Archives as Useful Primary Sources for Historians: Looking for Flame Wars." *Internet Histories* 2, no. 1-2 (2018): 38-54.

Hundschell, Andreas, Stefan Razinkas, Juia Backmann, and Martin Hoegl. "The Effects of Diversity on Creativity: A Literature Review and Synthesis." *Applied Psychology* 71 (2022): 1598-634.

Hyman, Avi. "Twenty Years of ListServ as an Academic Tool." *The Internet and Higher Education*, no. 6 (2003): 17-24.

Kirstein, Peter. "Early Experiences With the Arpanet and Internet in the United Kingdom." *IEEE Annals for the History of Computing* 21, no. 1 (1999): 38-44.

Knorr-Cetina, Karin. *Epistemic Cultures. How the Sciences Make Knowledge*. Harvard University Press, 1999.

Kolb, Rocky. "A Thousand Invisible Cords Binding Astronomy and High-Energy Physics." *Reports on Progress in Physics* 70, no. 10 (2007): 1583-95. https://doi.org/10.1088/0034-4885/70/10/R01.

Laidlaw, Emily B. "The Internet as a Democratising Force." In *Regulating Speech in Cyberspace: Gatekeepers, Human Rights and Corporate Responsibility*. Cambridge University Press, 2015.

Lewenstein, Bruce V. "Do Public Electronic Bulletin Boards Help Create Scientific Knowledge? The Cold Fusion Case." *Science, Technology & Human Values* 20, no. 2 (1995): 123-49.

Marra, Monica. "Arxiv-based commenting resources by and for astrophysicists and physicists: an initial survey." In *Expanding Perspectives on*

*Open Science: Communities, Cultures and Diversity in Concepts and Practices*, edited by Leslie Chan and Fernando Loizides. IoS Press, 2017. https://doi.org/10.3233/978-1-61499-769-6-100.

Matzat, Uwe. "Academic Communication and Internet Discussion Groups: Transfer of Information or Creation of Social Contacts?" *Social Networks* 26, no. 3 (2004): 221-55.

Mauldin, Michael L. "Empirical studies." In *Conceptual Information Retrieval. A Case Study in Adaptive Partial Parsing*, Springer, 1991.

Meadows, Arthur J. *Communicating research.* Academic Press, 1998.

Mieszkowski, Katharine. "The Geeks Who Saved Usenet." *Salon*, January 8, 2002, archived June 20, 2022, at https://web.archive.org/web/20220620153342/https://www.salon.com/2002/01/08/saving_usenet/.

Miller, Tristan, Camille Paloque-Bergès, and Avery Dame-Griff. "Remembering Netizens: An Interview with Ronda Hauben, Co-Author of Netizens: On the History and Impact of Usenet and the Internet (1997)." *Internet Histories* 7, no.1 (2023): 76-98. https://doi.org/10.1080/24701475.2022.2123120.

Milner, Helen V. "The Global Spread of the Internet: The Role of International Diffusion Pressures in Technology Adoption." Presented at the 2nd Conference on Interdependence, Diffusion, and Sovereignty, UCLA, California, March 2003. UCLA, 2003. https://ssrn.com/abstract=2182083.

Murtagh, Fionn. "Computer Networking in Astronomy." In *Information & On-Line Data in Astronomy,* edited by Daniel Egret and Miguel Albrecht. Springer, 1995.

Nordmann, Alfred, Hans Radder, and Gregor Schiemann, eds., *Science Transformed? Debating Claims of an Epochal Break* (University of Pittsburgh Press, 2011).

Paloque-Bergès, Camille. "Usenet as a Web Archive: Multi-Layered Archives of Computer-Mediated Communication." In *Web 25: Histories from the First 25 Years of the World Wide Web*, edited by Niels Brugger. Lang, 2017.

Paloque-Bergès, Camille. *Qu'est-Ce Qu'un Forum Internet? Une Généalogie Historique Au Prisme Des Cultures Savantes Numériques.* OpenEdition Press, 2018. https://doi.org/10.4000/books.oep.1843.

Pegoraro, Rob. "End of an Era: Google Groups to Drop Usenet Support." *PC Mag*, December 16, 2023, archived November 26, 2024, at

https://web.archive.org/web/20241126143729/https://www.pcmag.com/news/end-of-an-era-google-groups-to-drop-usenet-support.

Perkmann, Markus, Rossella Salandra, Valentina Tartari, Maureen McKelvey, and Alan Hughes. "Academic Engagement: A Review of the Literature 2011-2019." *Research Policy* 50, no. 1 (2021): 104114. https://www.sciencedirect.com/science/article/pii/S004873332030189X.

Peters, Benjamin. "On Digital Media History." *Journal of Communication* 74, no. 6 (2024): 509-12. https://doi.org/https://doi.org/10.1093/joc/jqae028.

Rosendhal, Jeffrey D., Philip Sakimoto, R. Pertzborn, and Larry Cooper. "The NASA Office of Space Science Education and Public Outreach Program." *Advances in Space Research* 34, no. 10 (2004): 2127-35.

Rowlands, Ian, David Nicholas, Bill Russell, Nicholas Canty, and Anthony Watkinson. "Social Media Use in the Research Workflow." *Learned Publishing* 24, no. 3 (2011): 183-95. https://doi.org/10.1087/20110306.

Schafer, Valérie. "Les Réseaux Sociaux Numériques d'avant…." *Le Temps Des Médias* 2, no. 31 (2018): 121-36. https://doi.org/10.3917/tdm.031.0121.

Smith, Marc A. "Invisible Crowds in Cyberspace: Mapping the Social Structure of the Usenet." In *Communities in Cyberspace*, edited by Marc A. Smith and Peter Kollock. Routledge, 1999. https://courses.ischool.berkeley.edu/i290-12/f06/smith_invisible_crowds.pdf.

Stebbins, Robert A. "Amateur and Professional Astronomers: A Study of Their Interrelationships." *Urban Life* 10, no. 4 (1982).

Stebbins, Robert A. "Amateurs and Their Place in Professional Science." In *New Generation Small Telescopes*, edited by D.S. Hayes, D.R. Genet, and R.M. Genet. Fairborn Press, 1987.

Stebbins, Robert A. "Looking Downwards - Sociological Images of the Vocation and Avocation of Astronomy." *Journal of the Royal Astronomical Society of Canada* 75 (1981): 2-14.

Strasser, Bruno J., Jérôme Baudry, Dana Mahr, Gabriela Sanchez, and Elise Tancoigne. "'Citizen Science'? Rethinking Science and Public Participation." *Science & Technology Studies* 32 no.2 (2019): 52-76. https://doi.org/10.23987/sts.60425.

Thagard, Paul. "Internet Epistemology: Contributions of New Information Technologies to Scientific Research." Copyright 1997. http://cogsci.uwaterloo.ca/Articles/Pages/Epistemology.html.

Traweek, Sharon. *Beamtimes and Lifetimes: The World of High Energy Physicists.* Harvard University Press, 1988.

Wellman, Barry, and Milena Gulia. "Virtual Communities as Communities: Net Surfers Don't Ride Alone." In *Communities in Cyberspace*, edited by Peter Kollock and Marc Smith. Routledge, 1999.

Williams, Thomas R. "Criteria for Identifying an Astronomer as an Amateur." In *Stargazers. The Contribution of Amateurs to Astronomy, Proceedings of Colloquium 98 of the IAU, June 20-24, 1987*, edited by Storm Dunlop and Michèle Gerbaldi. Springer, 1988.

The foundation of two very early Usenet newsgroups in astrophysics, still existent today, and some milestones in their history have been tracked from the origins at Princeton University in 1983 to 1994. They result to be pioneering experiences in this discipline, and among the earliest ones of this kind in academic disciplines at large. To the best of our knowledge, an account of their birth and evolution is given here for the first time, within the framework of some major steps in the evolution of Usenet newsgroups, one of the early Internet networks.
In compliance with key recommendations from the recent discipline of web history, this research has combined multiple and different type sources, building mainly on online archives of Usenet newsgroups and on human contributions from the concerned scholarly community. A final overview is proposed on how these early online communication tools have been perceived and used by the scholarly community involved.
This research reconstructs computer-mediated communication experiences which were at risk of being forgotten; provides a view of this environment's uptake of new communication technology; and contributes knowledge of some social dynamics of the astrophysics community in the last twenty years of the twentieth century.

Monica Marra earned a PhD in Italian Studies from the University of Pisa and, later on, a degree in Archivistics from the Archivio di Stato of Bologna. She has published studies and research about online scholarly communication in astrophysics. She works at the Italian National Institute for Astrophysics (INAF).

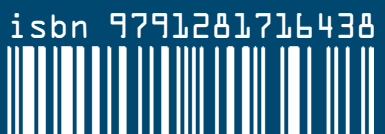